\documentclass[12pt,thmsa]{article}
\usepackage{amsfonts}
\usepackage{amsmath}
\usepackage{amssymb}
\usepackage{graphicx}
\usepackage{sw20mitp}

\input{tcilatex}
\renewcommand{\textwidth}{15cm}

\begin{document}

\author{Jean Petitot\thanks{%
\noindent CAMS, \'{E}cole des Hautes \'{E}tudes en Sciences Sociale, Paris. 
\newline
\indent\hspace{0.2cm} e-mail~: petitot@ehess.fr.fr }}
\title{Complexity and self-organization in Turing}
\date{}
\maketitle

\begin{center}
International Academy of Philosophy of Science Conference

The Legacy of A.M. Turing

Urbino, 25-28 september 2012
\end{center}

\bigskip

\section{Introduction: \textquotedblleft converting chemical information
into a geometrical form\textquotedblright}

Alan Turing's celebrated paper \textquotedblleft The Chemical Basis of
Morphogenesis\textquotedblright\ \cite{Turing52} published in 1952 in the 
\emph{Philosophical Transactions of the Royal Society of London} is a
typical example of a pioneering and inspired work in the domain of
mathematical modelling.

\begin{enumerate}
\item The paper presents a new key idea for solving an old problem. As
Lionel Harrison said in his 1987 paper \cite{Harrison87} \textquotedblleft
it is a theoretical preconception preceeding experience\textquotedblright .

\item It contains right away the germ of quite the whole theory associated
to the new ideas.

\item Its forsightedness is striking.\ It anticipates by many years its
experimental confirmations and mathematical developments.
\end{enumerate}

The key idea is formulated from the outset in the first sentence:

\begin{quotation}
\noindent \textquotedblleft It is suggested that a system of chimical
substances, called morphogens, reacting together and diffusing through a
tissue, is adequate to account for the main phenomena of
morphogenesis.\textquotedblright
\end{quotation}

\noindent Later (1953), Turing used a striking formulation (see below figure %
\ref{Inedit2}):

\begin{quotation}
\noindent \textquotedblleft It was suggested in Turing (1952) that this
might be the main means by which the chemical information contained in the
genes was converted into a geometrical form.\textquotedblright
\end{quotation}

Turing's colleague Claude Wilson Wardlaw, a botanist at the Department of
Cryptogamic Botany at Manchester University who wrote with Turing the 
paper \textquotedblleft A diffusion
reaction theory of morphogenesis in plants\textquotedblright\ \cite%
{Wardlaw52}, said in 1952 that Turing's working hypothesis was that:

\begin{quotation}
\noindent \textquotedblleft a localized accumulation of gene-determined
substances may be an essential prior condition [for cell
differentiation].\textquotedblright ~(p.~40)
\end{quotation}

\noindent and that what is needed in addition to biochemistry for
understanding emerging spatial forms in morphogenetic processes is a
\textquotedblleft patternized distribution of morphogenetic
substances\textquotedblright . There exist many homologies of organization
between different biological species and it must therefore exist general
morphogenetics mechanisms largely independent of specific genes.\ As he
claimed, \textquotedblleft certain physical processes are of very general
occurrence\textquotedblright ~(p.~46).

Wardlaw summarizes Turing's key idea in the following way:

\begin{quotation}
\noindent \textquotedblleft In an embryonic tissue in which the metabolic
substances may initially be distributed in a homogeneous manner, a regular,
patternized distribution of specific metabolites may eventually result, thus
affording the basis for the inception of morphological or histological
patterns.\textquotedblright ~(p.~44)
\end{quotation}

So, for understanding how a \emph{spatial} order can emerge from \emph{%
biochemical} reactions genetically controlled~-- which, according to Turing,
is the main problem of morphogenesis~-- Turing defines from the outset a
form as a \emph{breaking of homogeneity} of some spatially extended
biological tissue and, therefore, as a breaking of the symmetries underlying
such an homogeneity.

\section{The framework}

\subsection{Turing diffusion-driven instability}

The components of Turing's theorization are the following. Inside the
biological tissue under consideration, chimical reactions occur, which we
will call \emph{internal chimical dynamics}. In the spatial extension of
this substrate, processes of spatial diffusion occur, which we will call 
\emph{external spatial dynamics}. The key idea is that the \emph{coupling}
of these two very different kinds of dynamics can trigger, under certain
conditions, morphogenetic processes, which can be mathematically modelled by
using what are called since Turing \emph{reaction-diffusion differential
equations}. Why and how? Because the external spatial diffusion can \emph{%
destabilize}, under certain conditions, the internal chimical equilibria.

We must emphasize the fact that the notion of diffusion-driven instabilities
is to some extent paradoxical.\ Indeed, diffusion is a stabilizing process
and therefore the idea amounts to posit that the coupling of two stabilities
can induce an instability!

We will see that Turing assumes that there exists only one equilibrium of
the internal chimical dynamics. A diffusion-induced instability could
therefore make the chimical state diverge, but in general non-linearities of
the equations bound such divergences. Another possibility would be that
there exist several equilibria.\ Then Turing instabilities would induce
bifurcations from one equilibrium state to another one. It is this idea that
has been worked out in the late 1960s by Ren\'{e} Thom \cite{Thom72}, \cite%
{Thom74} to explain also morphogenesis.

\subsection{Turing's objective}

Today, reaction-diffusion equations are mainly used to explain the formation
of patterns in material substrates.\ But Turing's objective was deeper and
more ambitious and concerned embryogenesis.\ As he claimed in his
Introduction,

\begin{quotation}
\noindent \textquotedblleft The purpose of this paper is to discuss a
possible mechanism by which the genes of a zygote may determine the
anatomical structure of the resulting organism.\textquotedblright ~(p.~37)
\end{quotation}

As this general objective was too ambitious, he assumed many
simplifications. The first simplification was to eliminate any direct
reference to specific genes and to reduce the internal chemical dynamics to
reaction equations between concentrations of \emph{morphogens}. As explains
Philip Maini in \cite{Maini12}, a morphogen is

\begin{quotation}
\noindent \textquotedblleft a chemical to which cells respond by
differentiating in a concentration-dependent way.\textquotedblright
\end{quotation}

\noindent Turing was inspired by what Waddington \cite{Waddington40} called
\textquotedblleft form producers\textquotedblright\ or \textquotedblleft
evocators\textquotedblright . Morphogens are controlled by genes which
catalize their production, but, contrary to genes, they can diffuse in the
developing tissues and carry the \emph{positional information} (e.g. in the
sense of Wolpert) which is needed for morphogenesis.\footnote{%
For an introduction to the concept of \textquotedblleft positional
information\textquotedblright\ in Waddington, Wolpert, Goodwin and Thom, see
Petitot \cite{P03}.}

The second simplification made by Turing was to eliminate the
mechano-chemical aspects of embryogenesis, although he was aware of their
importance during the development. These aspects will be worked out later by
specialists such as George Oster and James Murray

But, even so drastically simplified, the search for good mathematical models
remains a

\begin{quotation}
\noindent \textquotedblleft a problem of formidable mathematical
complexity\textquotedblright ~(p.~38).
\end{quotation}

\begin{figure}[tbp]
\begin{center}
\includegraphics[hiresbb = true, width= 15cm,height= 5.48cm]{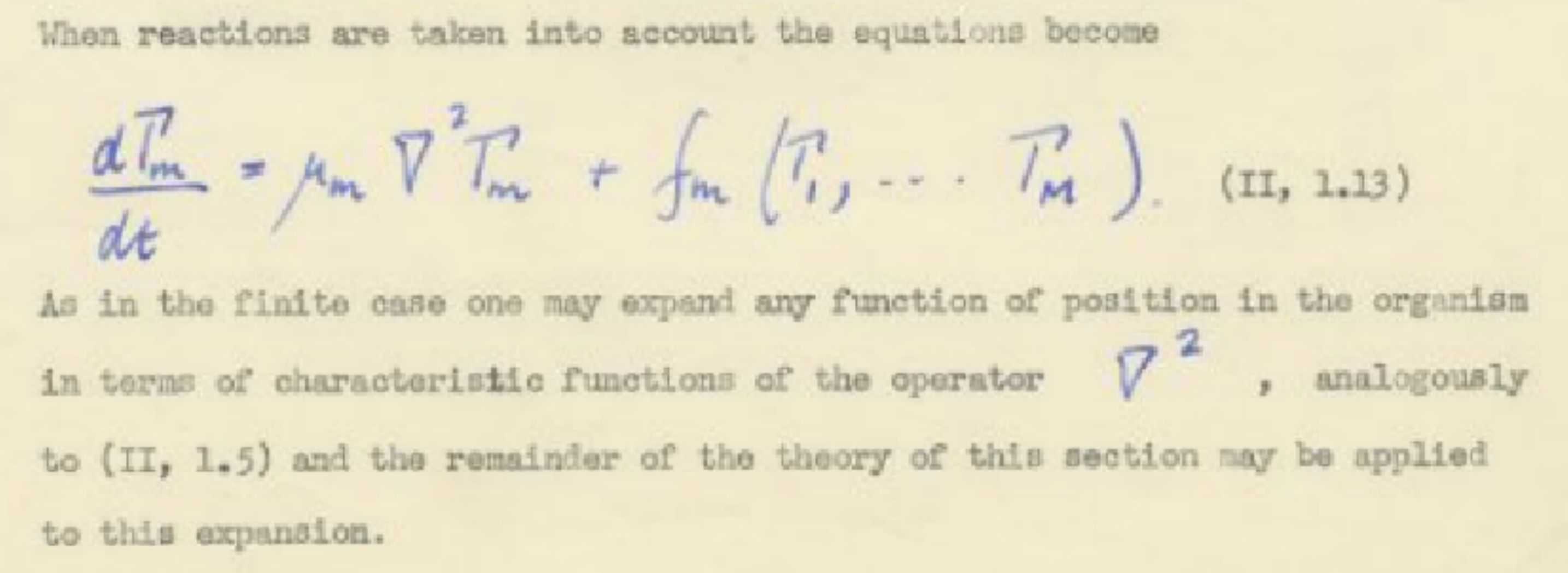}
\end{center}
\caption{Turing's general equation for morphogenesis.}
\label{Inedit1}
\end{figure}

In fact, it seems that Turing was looking for a kind of \emph{universal}
equation for morphogenesis.\ In his last paper \textquotedblleft Morphogen
theory of phyllotaxis\textquotedblright , which remained unpublished because
of his suicide and is kept at the King's College Archives, he proposed the
equation

\begin{equation*}
\frac{d\Gamma _{m}}{dt}=\mu _{m}\nabla ^{2}\Gamma _{m}+f_{m}\left( \Gamma
_{1},\cdots ,\Gamma _{M}\right)
\end{equation*}

\noindent where the $\Gamma _{m}$ are the respective concentrations of the $%
M $ morphogens, $\nabla ^{2}$ the spatial Laplacian, that is the diffusion
operator, the $\mu _{m}$ the diffusibility coefficients, and the $f_{m}$ the
reaction equations (see figure \ref{Inedit1})



\begin{figure}[tbp]
\begin{center}
\includegraphics[
width= 15cm,height= 4.71cm]{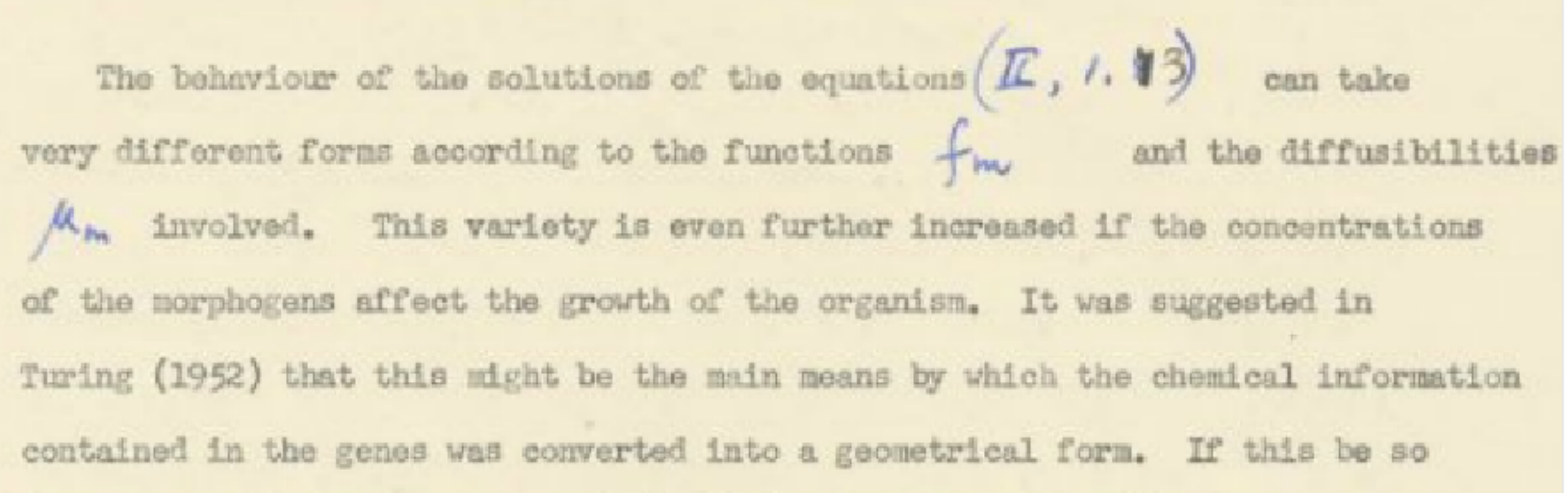}
\end{center}
\caption{Turing's anticipation of the richness of the general equation for
morphogenesis.}
\label{Inedit2}
\end{figure}

The point was that, when you vary the $f_{m}$ and the $\mu _{m}$, the
solutions of such a universal equation can be extremely diverse (see figure %
\ref{Inedit2}). Hence the idea that, as with Newton's equation for
Mechanics, it could be possible to classify a lot of very different kinds of
forms using the same general equation.



Indeed, we can vary three classes of parameters:

\begin{enumerate}
\item For the chemical part, the eigenvalues provided by the spectral
analysis of the linearized system of the $f_{m}$.

\item The diffusibility coefficients $\mu _{m}$.

\item For the geometrical part, the eigenfunctions of the Laplacian operator
(harmonic analysis).
\end{enumerate}

\subsection{Turing's foresightedness}

In his 1990 paper \textquotedblleft Turing's theory of morphogenesis.\ Its
influence on modelling biological pattern and form\textquotedblright\ \cite%
{Murray90}, James Murray claims that Turing's 1952 paper is
\textquotedblleft one of the most important papers in theoretical biology of
this century\textquotedblright ~(p.~119). Indeed,

\begin{quotation}
\noindent \textquotedblleft What is astonishing about Turing's seminal paper
is that, with very few exceptions, it took the mathematical world more than
20 years to realise the wealth of fascinating problems posed by his theory.
What is even more astonishing is that it was closer to 30 years before a
significant number of experimental biologists took serious notice of its
implications and potential applications in developmental biology, ecology
and epidemiology.\textquotedblright ~(p.~121)
\end{quotation}

These inspired anticipations proved to be exact in chemistry.\ There exist
today a lot of models of chimical reaction-diffusion phenomena: clocks,
travelling waves, etc. Their analysis constitutes a rapidly expanding
research domain while, at Turing's time, no empirical example was known.
Turing discovered theoretically the basic phenomenon and was the first to
compute simulations on the computer he had himself constructed at
Manchester. In embryogenesis, the exact limits of validity of Turing model
are still under discussion.

\section{The context}

\subsection{The bibliography}

It is interesting to look at Turing's bibliography, which is very short.\
First, it includes two books which are not really used, the \emph{Theory of
Elasticity and Magnetism} of James Jeans (1927) and \emph{The permeability
of natural membranes} of Hugh Dawson and James Danielli (1943). Then, it
cites a fundamental paper of Leonor Michaelis and Maud Menten (1913) on 
\emph{Die Kinetik der Invertinwirkung} \cite{MichLent} whose pioneering
mathematical model is typical of the internal chimical dynamics used by
Turing. Finally, there are three masterpieces on embryology and
morphogenesis: Charles Manning Child's \textquotedblleft
summa\textquotedblright\ \emph{Patterns and problems of development} (1941),
Sir D'Arcy Thompson's masterpiece \emph{On Growth and Form} (1942) \cite%
{Arcy} and Conrad Hal Waddington's key work on \emph{Organizers and Genes}
(1940) \cite{Waddington40}. Introduced by Hans Spemann, \textquotedblleft
organizers\textquotedblright\ were thought to be the cause of the
embryological induction observed when the tissues of some part of an embryo
(e.g. a leg) were transplanted in another part (e.g. the head). The idea
(much speculative at that time) was that there must exist chemical signals
triggering cellular differenciations.\ It is in the second part of this work
that Waddington assumed that, through morphogens, gene concentrations could
be important for cellular differenciation and that the developmental units
of an organism are \textquotedblleft morphogenetic fields\textquotedblright .

\subsection{The kinetic model}

Building on previous very precise numerical experimental data collected by
Victor Henri (1903), the Michaelis-Menten model of the kinetics of invertase
enzyme (1913) was the first to explain the catalysis of the hydrolysis of
sucrose into glucose and fructose.\ Let $E$ be an enzyme bounding with a
substrate $S$ to give a complex $ES$ which converts itself into a product $P$
through a chain of two elementary chemical reactions:

\begin{equation*}
E+S\overset{k_{1}}{\underset{k_{2}}{\rightleftarrows }}ES\overset{k_{3}}{%
\rightarrow }E+P
\end{equation*}

\noindent where the $k_{i}$ are the rate constants of the reactions. Let us
denote by $\left[ X\right] $ the concentration of $X$.\ Then the law of mass
action saying that a reaction rate is proportional to the product of the
concentrations of the reactants implies the system of nonlinear differential
equations:\footnote{$\dot{X}$ is the traditional notation for the temporal
derivative $\frac{dX}{dt}$.}

\begin{equation*}
\left\{ 
\begin{array}{l}
\overset{\cdot }{\left[ S\right] }=-k_{1}\left[ E\right] \left[ S\right]
+k_{2}\left[ ES\right] \\ 
\overset{\cdot }{\left[ E\right] }=-k_{1}\left[ E\right] \left[ S\right]
+k_{2}\left[ ES\right] +k_{3}\left[ ES\right] \\ 
\overset{\cdot }{\left[ ES\right] }=k_{1}\left[ E\right] \left[ S\right]
-k_{2}\left[ ES\right] -k_{3}\left[ ES\right] \\ 
\overset{\cdot }{\left[ P\right] }=k_{3}\left[ ES\right]%
\end{array}%
\right.
\end{equation*}

\noindent where the relation $\overset{\cdot }{\left[ E\right] }+\overset{%
\cdot }{\left[ ES\right] }=0$ implies the conservation law $\left[ E\right] +%
\left[ ES\right] =E_{0}=$ constant. Under an hypothesis of adiabaticity
according to which the equilibrium between $S$ and $ES$ is \textquotedblleft
instantaneous\textquotedblright , that is $\overset{\cdot }{\left[ S\right] }%
=0$, then $k_{1}\left[ E\right] \left[ S\right] =k_{2}\left[ ES\right] $, $%
\left[ ES\right] =\frac{k_{1}}{k_{2}}\left[ E\right] \left[ S\right] =\frac{%
k_{1}}{k_{2}}\left[ S\right] \left( E_{0}-\left[ ES\right] \right) $, $\left[
ES\right] \left( 1+\frac{k_{1}}{k_{2}}\left[ S\right] \right) =\frac{k_{1}}{%
k_{2}}\left[ S\right] E_{0}$ and 
\begin{equation*}
\left[ ES\right] =\frac{k_{1}}{k_{2}}\left[ S\right] E_{0}\left( \frac{1}{%
\left( 1+\frac{k_{1}}{k_{2}}\left[ S\right] \right) }\right) =\left[ S\right]
E_{0}\frac{1}{K}\left( \frac{1}{\left( 1+\frac{1}{K}\left[ S\right] \right) }%
\right) =\frac{E_{0}\left[ S\right] }{K+\left[ S\right] }
\end{equation*}%
with $K=\frac{k_{2}}{k_{1}}$, and therefore

\begin{equation*}
\overset{\cdot }{\left[ P\right] }=k_{3}\frac{E_{0}\left[ S\right] }{K+\left[
S\right] }
\end{equation*}

\section{Turing's numerical example}

So, Turing start with morphogens diffusing and reacting inside a tissue.\
Diffusion flows from regions of strong concentrations towards regions of
weak concentrations with a velocity proportional to the gradients of the
concentrations and to the diffusibility coefficients. According to the law
of mass action, reaction rates are proportional to the product of
concentrations.\ Hence a huge variety of nonlinear differential equations.

Turing gives several examples and develop one of them in minute detail in
his \S 10 \textquotedblleft A numerical example\textquotedblright . He
considers a ring of $N=20$ cells and two morphogens $X$ and $Y$ and makes
several numerical assumptions on their size, the diffusibility constants,
the permeability of membranes (it is here that the reference to
Dawson-Danielli is used), etc. With great acuity, he argues that the system
must be (thermodynamically) \emph{open} and include a \textquotedblleft fuel
substance\textquotedblright\ $A$ providing it with energy through its
degradation into another substance $B$.

\begin{quotation}
\noindent \textquotedblleft In order to maintain the wave pattern a
continual supply of free energy is required. It is clear that this must be
so since there is a continual degradation of energy through diffusion. This
energy is supplied through the `fuel substances' ($A$, $B$ in the last
example), which are degraded into `waste products'.\textquotedblright
~(p.~65)
\end{quotation}

\noindent For modelling catalysis, Turing add three other substances $C$, $%
C^{\prime }$, and $W$.

It must be emphasized that it will be only thirty years later that open
thermodynamical systems out of equilibrium will be systematically
investigated (see e.g. Prigogine's dissipative structures).\ 

\subsection{The internal chimical dynamics}

The system of $7$ elementary reactions proposed by Turing is the following:

\begin{equation*}
\begin{array}{llccl}
\left[ 1\right] & Y+X\rightarrow W &  &  & \text{rate: }\frac{25}{16}XY \\ 
\left[ 2\right] & W+A\rightarrow 2Y+B &  &  & 
\begin{array}{l}
\text{(instantly)} \\ 
A=1000=\text{constant (fuel substance)}%
\end{array}
\\ 
\left[ 3\right] & 2X\rightarrow W &  &  & \text{rate: }\frac{7}{64}X^{2} \\ 
\left[ 4\right] & A\rightarrow X &  &  & \text{rate: }\frac{1}{16}10^{-3}A=%
\frac{1}{16} \\ 
\left[ 5\right] & Y\rightarrow B &  &  & \text{rate: }\frac{1}{16}Y \\ 
\left[ 6\right] & Y+C\rightarrow C^{\prime } &  &  & 
\begin{array}{l}
\text{(instantly)} \\ 
C=C^{\prime }=10^{-3}\left( 1+\gamma \right)%
\end{array}
\\ 
\left[ 7\right] & C^{\prime }\rightarrow X+C &  &  & \text{rate: }\frac{55}{%
32}10^{3}C^{\prime }=\frac{55}{32}\left( 1+\gamma \right)%
\end{array}%
\end{equation*}

So, $X$ converts into $Y$ at the rate $\frac{1}{32}\left[ 50XY+7X^{2}-55%
\left( 1+\gamma \right) \right] $ (because of $\left[ 1\right] $, $\left[ 3%
\right] $, and $\left[ 7\right] $) while self-reproducing (because of $\left[
4\right] $) at the constant rate $\frac{1}{16}$ and destroying $Y$ (because
of $\left[ 5\right] $) at the rate $\frac{1}{16}Y$. So the kinetic equations
for the time varying concentrations $X\left( t\right) $, $Y\left( t\right) $
of the morphogens $X$, $Y$~-- i.e. the internal chimical dynamics~-- are

\begin{equation*}
\left\{ 
\begin{array}{l}
\overset{\cdot }{X}=\frac{1}{32}\left[ -50XY-7X^{2}+57+55\gamma \right]
=f\left( X,Y\right) \\ 
\overset{\cdot }{Y}=\frac{1}{32}\left[ 50XY+7X^{2}-55-55\gamma -2Y\right]
=g\left( X,Y\right)%
\end{array}%
\right.
\end{equation*}

\noindent where $57$ in the first equation is $55+2$ with $2$ coming from $%
\left[ 4\right] $ and $-2Y$ in the second equation comes from $\left[ 5%
\right] $.

\subsection{Equilibria and linearization}

A chemical internal \emph{equilibrium} corresponds to values $\left(
X_{e},Y_{e}\right) $ such that $f\left( X_{e},Y_{e}\right) =0$ and $g\left(
X_{e},Y_{e}\right) =0$.\ For $\gamma =0$ the system is

\begin{equation*}
\left\{ 
\begin{array}{l}
\overset{\cdot }{X}=\frac{1}{32}\left[ -50XY-7X^{2}+57\right] =f\left(
X,Y\right) \\ 
\overset{\cdot }{Y}=\frac{1}{32}\left[ 50XY+7X^{2}-55-2Y\right] =g\left(
X,Y\right)%
\end{array}%
\right.
\end{equation*}

\noindent and an evident equilibrium is $X_{0}=Y_{0}=1$.\ Another is $X_{1}=-%
\frac{57}{7},Y_{1}=1$ but it is non physical since a concentration cannot be
negative.

Then, Turing linearizes the system near the equilibrium $\left(
X_{0},Y_{0}\right) $ and analyzes the stability of the linear system. The
method was already well known at his time: the matrix of the linearized
system is the Jacobian $J_{0}$ of $\left\{ f,g\right\} $ at $\left(
X_{0},Y_{0}\right) $ and the stability depends upon the fact that the \emph{%
real part} of all eigenvalues $\lambda $ of $J_{0}$ are $<0$. Let us briefly
remind non mathematicians of it. It is immediate to compute $J_{0}$ for $%
\gamma =0$:

\begin{equation*}
J_{0}=\left( 
\begin{array}{cc}
\frac{\partial f}{\partial X} & \frac{\partial f}{\partial Y} \\ 
\frac{\partial g}{\partial X} & \frac{\partial g}{\partial Y}%
\end{array}%
\right)
\end{equation*}

\noindent and, as

\begin{equation*}
\left\{ 
\begin{array}{l}
\frac{\partial f}{\partial X}=a=\frac{1}{32}\left[ -50Y_{0}-14X_{0}\right]
=-2 \\ 
\frac{\partial f}{\partial Y}=b=\frac{1}{32}\left[ -50X_{0}\right] =-\frac{25%
}{16}=-1.5625 \\ 
\frac{\partial g}{\partial X}=c=\frac{1}{32}\left[ 50Y_{0}+14X_{0}\right] =2
\\ 
\frac{\partial g}{\partial Y}=d=\frac{1}{32}\left[ 50X_{0}-2\right] =\frac{48%
}{32}=\frac{3}{2}=1.5%
\end{array}%
\right.
\end{equation*}

\noindent we get

\begin{equation*}
J_{0}=\left( 
\begin{array}{cc}
-2 & -\frac{25}{16} \\ 
2 & \frac{3}{2}%
\end{array}%
\right)
\end{equation*}

In a small neighbourhood of $\left( X_{0},Y_{0}\right) $ we can write $%
X=X_{0}+x$, $Y=Y_{0}+y$ and write at first order

\begin{equation*}
\left( 
\begin{array}{c}
\dot{x} \\ 
\dot{y}%
\end{array}%
\right) =J_{0}\left( 
\begin{array}{c}
x \\ 
y%
\end{array}%
\right) ,~\text{i.e. }\left\{ 
\begin{array}{l}
\dot{x}=ax+by \\ 
\dot{y}=cx+dy%
\end{array}%
\right.
\end{equation*}

\noindent As the system is linear, we look at solutions of the form

\begin{equation*}
\left( 
\begin{array}{c}
x\left( t\right) \\ 
y\left( t\right)%
\end{array}%
\right) =e^{\lambda t}\left( 
\begin{array}{c}
x_{0} \\ 
y_{0}%
\end{array}%
\right)
\end{equation*}

\noindent where $\left( x_{0},y_{0}\right) $ is the state of the system at
time $t=0$. They are straight trajectories on the line $\left( 0,0\right)
-\left( x_{0},y_{0}\right) $ with an exponential temporal law. Computing the
derivatives in two different ways, we get

\begin{equation*}
\left( 
\begin{array}{c}
\dot{x} \\ 
\dot{y}%
\end{array}%
\right) =J_{0}\left( 
\begin{array}{c}
x \\ 
y%
\end{array}%
\right) =\lambda e^{\lambda t}\left( 
\begin{array}{c}
x_{0} \\ 
y_{0}%
\end{array}%
\right) =\lambda \left( 
\begin{array}{c}
x \\ 
y%
\end{array}%
\right)
\end{equation*}

\noindent that is an equation linking $\lambda $ to $J_{0}$:

\begin{equation*}
\left( J_{0}-\lambda I\right) \left( 
\begin{array}{c}
x \\ 
y%
\end{array}%
\right) =0
\end{equation*}

\noindent If $\left( x_{0},y_{0}\right) \neq \left( 0,0\right) $, then $%
\left( x,y\right) \neq \left( 0,0\right) $ and this equation can be
satisfied only if the \emph{determinant} $\func{Det}\left( J_{0}-\lambda
I\right) $ vanishes.\ The equation $\func{Det}\left( J_{0}-\lambda I\right)
=0$ is called the \emph{characteristic equation} of the linear system. It is
a polynomial equation of degree $2$ which writes

\begin{eqnarray*}
\func{Det}\left( 
\begin{array}{cc}
a-\lambda & b \\ 
c & d-\lambda%
\end{array}%
\right) &=&0 \\
\lambda ^{2}-\left( a+d\right) \lambda +ad-bc &=&0 \\
\lambda ^{2}-\limfunc{Tr}\left( J_{0}\right) \lambda +\func{Det}\left(
J_{0}\right) &=&0 \\
\lambda ^{2}-S\lambda +P &=&0
\end{eqnarray*}

\noindent where the sum of diagonal terms $\limfunc{Tr}\left( J_{0}\right)
=a+d=S$, called the \emph{trace} of the matrix $J_{0}$, gives the sum $S$ of
the solutions and the determinant $\func{Det}\left( J_{0}\right) $ of $J_{0}$
gives their product $P$. As the discriminant of the equation is $\Delta
=S^{2}-4P=\limfunc{Tr}\left( J_{0}\right) ^{2}-4\func{Det}\left(
J_{0}\right) $, the solutions are given by the well known formula

\begin{eqnarray*}
\lambda _{\pm } &=&\frac{1}{2}\left( S\pm \sqrt{\Delta }\right) \\
\lambda _{\pm } &=&\frac{1}{2}\left( \limfunc{Tr}\left( J_{0}\right) \pm 
\sqrt{\limfunc{Tr}\left( J_{0}\right) ^{2}-4\func{Det}\left( J_{0}\right) }%
\right)
\end{eqnarray*}

\noindent and any solution of the linear system is a linear combination of
the two solutions with $\lambda _{\pm }$.

Let us suppose now that $\lambda =\alpha +i\omega $ has real part $\limfunc{%
Re}\left( \lambda \right) =\alpha $ and imaginary part $\func{Im}\left(
\lambda \right) =\omega $.\ Then, $e^{\lambda t}=e^{\left( \alpha +i\omega
\right) t}=e^{\alpha t}e^{i\omega t}$ is an oscillation modulated by the
real exponential $e^{\alpha t}$.\ If $\alpha >0$, $e^{\alpha t}$ diverges
exponentially when $t\rightarrow +\infty $ and the corresponding
trajectories go to infinity and are unstable.\ On the contrary, if $\alpha
<0 $, $e^{\alpha t}$ converges exponentially towards $0$ when $t\rightarrow
+\infty $ and the corresponding trajectories go to equilibrium and are
stable.\ In Turing's example,

\begin{eqnarray*}
\limfunc{Tr}\left( J_{0}\right) &=&a+d=-2+\frac{3}{2}=-\frac{1}{2} \\
\func{Det}\left( J_{0}\right) &=&ad-bc=-2\times \frac{3}{2}-\left( -\frac{25%
}{16}\right) \times 2=\frac{1}{8} \\
\lambda _{\pm } &=&-\frac{1}{4}\left( 1\pm i\right) ,~\limfunc{Re}\left(
\lambda _{\pm }\right) =-\frac{1}{4}<0
\end{eqnarray*}

\noindent and the two eigenvalues have negative real parts.\ The internal
chemical equilibrium $\left( X_{0},Y_{0}\right) $ (i.e. $\left(
x_{0},y_{0}\right) =0$) is therefore \emph{stable}.

More generally, in the $2$-dimensional case, the system is stable if and
only if

\begin{equation*}
\left\{ 
\begin{array}{c}
\limfunc{Tr}\left( J_{0}\right) =a+d<0,~\text{here }-\frac{1}{2}<0 \\ 
\func{Det}\left( J_{0}\right) =ad-bc>0,~\text{here }\frac{1}{8}>0%
\end{array}%
\right.
\end{equation*}

\noindent Indeed, we must have $\limfunc{Re}\left( \lambda _{\pm }\right) <0$%
. If $\Delta =\limfunc{Tr}\left( J_{0}\right) ^{2}-4\func{Det}\left(
J_{0}\right) <0$, then $\lambda _{+}$ and $\lambda _{-}$ are complex
conjugate eigenvalues and we must have $\limfunc{Tr}\left( J_{0}\right) <0$,
and of course $\func{Det}\left( J_{0}\right) >0$ because otherwise we would
have $\Delta >0$.\ If $\Delta \geq 0$, then $\lambda _{+}$ and $\lambda _{-}$
are real eigenvalues and they must be both $<0$.\ This imply that the
greatest eigenvalue, namely $\lambda _{+}=\limfunc{Tr}\left( J_{0}\right) +%
\sqrt{\Delta }$ must be $<0$, which implies $\limfunc{Tr}\left( J_{0}\right)
<-\sqrt{\Delta }$.\ So $\limfunc{Tr}\left( J_{0}\right) <0$, and, as $%
\limfunc{Tr}\left( J_{0}\right) ^{2}>\Delta =\limfunc{Tr}\left( J_{0}\right)
^{2}-4\func{Det}\left( J_{0}\right) $, we have also $\func{Det}\left(
J_{0}\right) >0$.

\section{Diffusion-driven instability}

After having defined the internal chemical equilibrium and analyzed its
stability, Turing explains how a spatial diffusion of the morphogens $X$, $Y$
can induce an \emph{instability}. Let us summarize his computations.

\subsection{The reaction-diffusion model}

Let $r=1,\cdots ,N$ label the positions of the $N$ cells in the ring.
Concentrations $X$, $Y$ are then functions $X(r,t)$ and $Y(r,t)$ of time $t$
and spatial position $r$. In a continuous model, the spatial positions would
be parametrized by an angle $\theta \in \mathbb{S}^{1}$ and concentrations
would be functions $X(\theta ,t)$ and $Y(\theta ,t)$ (angles $\theta
_{r}=2\pi \frac{r}{N}$ retrieve the discrete case). We start with an \emph{%
homogeneous} initial state where $X(r,t)=X_{0}$ and $Y(r,t)=Y_{0}$
everywhere and we apply diffusion using the Laplace operator $\Delta =\nabla
^{2}$ and its discrete approximation $\Delta F\left( r\right) =F\left(
r-1\right) -2F\left( r\right) +F\left( r+1\right) $ for any function $%
F\left( r\right) $.

We get that way the \emph{reaction-diffusion} equations

\begin{equation*}
\left\{ 
\begin{array}{l}
\overset{\cdot }{X}(r,t)=f\left( X(r,t),Y(r,t)\right) + \\ 
\hspace{1cm}\mu \left( X(r-1,t)-2X(r,t)+X(r+1,t)\right) \\ 
\overset{\cdot }{Y}(r,t)=g\left( X(r,t),Y(r,t)\right) + \\ 
\hspace{1cm}\nu \left( Y(r-1,t)-2Y(r,t)+Y(r+1,t)\right)%
\end{array}%
\right. ,~\left( r=1,\ldots ,N\right)
\end{equation*}

\noindent where $\mu $ and $\nu $ are the respective coefficients of
diffusibility of $X$ and $Y$. Some technical aspects of the general analysis
of such equations are well emphasized by Turing in \S 11 \textquotedblleft
Restatement and biological interpretation of the results\textquotedblright
~(p.~66), with an incredible sense of anticipation: it is essential to take
into account

\begin{enumerate}
\item the role of \emph{fluctuations}, which play a critical role when the
system becomes unstable;

\item the role of \emph{slow changes}\ of reaction rates and diffusibility
coefficients because \textquotedblleft such changes are supposed ultimately
to bring the system out of the stable state\textquotedblright .
\end{enumerate}

Turing considered therefore that the systems he analyzed belong to the class
of what are called today \emph{slow-fast dynamical systems} and focused on
the breaking of spatial homogeneity near instability.\ As he said

\begin{quotation}
\noindent \textquotedblleft the phenomena when the system is just unstable
were the particular subject of the inquiry.\textquotedblright
\end{quotation}

\noindent He underlined the fact that the \textquotedblleft linearity
assumption\textquotedblright\ near the equilibrium, i.e. the fact that the
dynamics is qualitatively equivalent to its linear part, is
\textquotedblleft a serious one\textquotedblright\ and made what is called
today an hypothesis of \emph{adiabaticity}: as the system is a slow-fast
one, the slow variation of parameters is slow w.r.t. the fast time used to
reach equilibrium and, therefore, one can suppose that the system is always
in its equilibrium state until he reaches a bifurcation destabilizing it.

In terms of the variables $x\left( r,t\right) $ and $y\left( r,t\right) $,
the linearized reaction-diffusion equations are:

\begin{equation*}
\left\{ 
\begin{array}{l}
\overset{\cdot }{x}(r,t)=ax(r,t)+by(r,t)+ \\ 
\hspace{1cm}\mu \left( x(r-1,t)-2x(r,t)+x(r+1,t)\right) \\ 
\overset{\cdot }{y}(r,t)=cx(r,t)+dy(r,t)+ \\ 
\hspace{1cm}\nu \left( y(r-1,t)-2y(r,t)+y(r+1,t)\right)%
\end{array}%
\right. ,~\left( r=1,\ldots ,N\right)
\end{equation*}

\noindent In the continuous limit on a circle of radius $1$, they are

\begin{equation*}
\left( 
\begin{array}{c}
\dot{x}\left( \theta ,t\right) \\ 
\dot{y}\left( \theta ,t\right)%
\end{array}%
\right) =J_{0}\left( 
\begin{array}{c}
x\left( \theta ,t\right) \\ 
y\left( \theta ,t\right)%
\end{array}%
\right) +\left( 
\begin{array}{cc}
\mu ^{\prime } & 0 \\ 
0 & \nu ^{\prime }%
\end{array}%
\right) \left( 
\begin{array}{c}
x^{\prime \prime }\left( \theta ,t\right) \\ 
y^{\prime \prime }\left( \theta ,t\right)%
\end{array}%
\right)
\end{equation*}

\noindent where $x^{\prime \prime }$ and $y^{\prime \prime }$ are spatial
second derivatives (Laplacian term).

A pedagogical interest of the ring model is that the space is the circle $%
\mathbb{S}^{1}$, that the eigenfunctions of the Laplacian are the
trigonometric functions, and that the harmonic analysis is therefore nothing
else than Fourier analysis. Let $\xi \left( s,t\right) $ and $\eta \left(
s,t\right) $ be the Fourier tranforms of $x\left( r,t\right) $ and $y\left(
r,t\right) $:

\begin{equation*}
\left\{ 
\begin{array}{l}
\xi \left( s,t\right) =\frac{1}{N}\sum_{r=1}^{r=N}\exp \left( -\frac{2\pi irs%
}{N}\right) x(r,t) \\ 
\eta \left( s,t\right) =\frac{1}{N}\sum_{r=1}^{r=N}\exp \left( -\frac{2\pi
irs}{N}\right) y(r,t)%
\end{array}%
\right. ,~\left( r=1,\ldots ,N\right)
\end{equation*}

\noindent Then $x\left( r,t\right) $ and $y\left( r,t\right) $ are retrieved
through the inverse Fourier transform:

\begin{equation*}
\left\{ 
\begin{array}{l}
x\left( r,t\right) =\sum_{s=1}^{s=N}\exp \left( \frac{2\pi irs}{N}\right)
\xi \left( s,t\right) \\ 
y\left( r,t\right) =\sum_{s=1}^{s=N}\exp \left( \frac{2\pi irs}{N}\right)
\eta \left( s,t\right)%
\end{array}%
\right. ,~\left( s=1,\ldots ,N\right)
\end{equation*}

\noindent By definition, the $\xi \left( s,t\right) $ and $\eta \left(
s,t\right) $ are complex numbers.\ But as far as $x\left( r,t\right) $ and $%
y\left( r,t\right) $ are real, we must have $\xi \left( s,t\right) =%
\overline{\xi \left( N-s,t\right) }$ and $\eta \left( s,t\right) =\overline{%
\eta \left( N-s,t\right) }$.

The main interest of using harmonic analysis, is that, in the Fourier
domain, the system of equations becomes \emph{diagonal} because the
functions are expanded over a basis of \emph{eigenfunctions} of the
Laplacian operator. Turing based his computations on this \emph{separation
of variables} in the Fourier domain. Due to the definition of $\xi \left(
s,t\right) $ and the expression of $\overset{\cdot }{x}(r,t)$, the temporal
derivatives $\overset{\cdot }{\xi }\left( s,t\right) $ are

\begin{eqnarray*}
\overset{\cdot }{\xi }\left( s,t\right) &=&\frac{1}{N}\sum_{r=1}^{r=N}\exp
\left( -\frac{2\pi irs}{N}\right) \\
&&\left[ ax(r,t)+by(r,t)+\mu \left( x(r-1,t)-2x(r,t)+x(r+1,t)\right) \right]
\end{eqnarray*}

\noindent If one writes $rs=\left( r+1\right) s-s$ and uses the
orthogonality relations between the eigenfunctions

\begin{eqnarray*}
\sum_{s=1}^{s=N}\exp \left( \frac{2\pi irs}{N}\right) &=&0\text{ if }%
r=1,\ldots ,N-1 \\
\sum_{s=1}^{s=N}\exp \left( \frac{2\pi irs}{N}\right) &=&N\text{ if }r=N
\end{eqnarray*}

\noindent then, one gets the equations

\begin{eqnarray*}
\overset{\cdot }{\xi }\left( s,t\right) &=&a\xi \left( s,t\right) +b\eta
\left( s,t\right) + \\
&&\mu \left( \exp \left( -\frac{2\pi is}{N}\right) -2+\exp \left( \frac{2\pi
is}{N}\right) \right)
\end{eqnarray*}

\noindent and analog formulae for the $\eta \left( s,t\right) $.\ One then
takes the real and imaginary parts of the equations and uses the formulae

\begin{eqnarray*}
\exp \left( \frac{2\pi is}{N}\right) &=&\cos \left( \frac{2\pi s}{N}\right)
+i\sin \left( \frac{2\pi s}{N}\right) \\
\sin \left( -\frac{2\pi s}{N}\right) +\sin \left( \frac{2\pi s}{N}\right)
&=&0 \\
\cos \left( -\frac{2\pi s}{N}\right) -2+\cos \left( \frac{2\pi s}{N}\right)
&=&2\left( \cos \left( \frac{2\pi s}{N}\right) -1\right) =2\left( \cos
^{2}\left( \frac{\pi s}{N}\right) -\sin ^{2}\left( \frac{\pi s}{N}\right)
-1\right) \\
&=&-4\sin ^{2}\left( \frac{\pi s}{N}\right) \text{ since }\cos ^{2}\left( 
\frac{\pi s}{N}\right) +\sin ^{2}\left( \frac{\pi s}{N}\right) =1
\end{eqnarray*}

\noindent to get the equations

\begin{equation*}
\left\{ 
\begin{array}{l}
\overset{\cdot }{\xi }\left( s,t\right) =\left( a-4\mu \sin ^{2}\left( \frac{%
\pi s}{N}\right) \right) \xi \left( s,t\right) +b\eta \left( s,t\right) \\ 
\overset{\cdot }{\eta }\left( s,t\right) =c\xi \left( s,t\right) +\left(
d-4\nu \sin ^{2}\left( \frac{\pi s}{N}\right) \right) \eta \left( s,t\right)%
\end{array}%
\right.
\end{equation*}

In the continuous model, the Fourier transforms of functions on $\mathbb{S}%
^{1}$ are Fourier series whose components are indexed by $k\in \mathbb{Z}$
and one gets

\begin{equation*}
\left\{ 
\begin{array}{c}
\overset{\cdot }{\xi }\left( k,t\right) =\left( a-\mu ^{\prime }k^{2}\right)
\xi \left( k,t\right) +b\eta \left( k,t\right) \\ 
\overset{\cdot }{\eta }\left( k,t\right) =c\xi \left( k,t\right) +\left(
d-\nu ^{\prime }k^{2}\right) \eta \left( k,t\right)%
\end{array}%
\right.
\end{equation*}

\noindent with $k^{2}$ corresponding to $\frac{N^{2}}{\pi ^{2}}\sin
^{2}\left( \frac{\pi s}{N}\right) $. Turing denotes by $U$ this variable.

\subsection{The origin of instability}

The fundamental new phenomenon introduced by diffusion is that the spectral
analysis of the linearized system now depends upon diffusion which, by
changing the characteristic equation, can tranform eigenvalues with $%
\limfunc{Re}\left( \lambda \right) <0$ into eigenvalues with $\limfunc{Re}%
\left( \lambda \right) >0$. It is the origin of diffusion-driven
instabilities. Indeed, the Jacobian is now 
\begin{equation*}
J=\left( 
\begin{array}{cc}
a-4\mu \sin ^{2}\left( \frac{\pi s}{N}\right) & b \\ 
c & d-4\nu \sin ^{2}\left( \frac{\pi s}{N}\right)%
\end{array}%
\right)
\end{equation*}

\noindent and the characteristic equation~-- also called a \emph{dispersion
relation}~ -- is therefore

\begin{equation*}
\left( p-a+4\mu \sin ^{2}\left( \frac{\pi s}{N}\right) \right) \left(
p-d+4\nu \sin ^{2}\left( \frac{\pi s}{N}\right) \right) =bc
\end{equation*}

Turing denotes by $p_{s}$ and $p_{s}^{\prime }$, with $\limfunc{Re}\left(
p_{s}\right) \geq \limfunc{Re}\left( p_{s}^{\prime }\right) $ the two
eigenvalues.\ If $p_{s}\neq p_{s}^{\prime }$ then the solutions of the
system in the Fourier domain are of the form

\begin{equation*}
\left\{ 
\begin{array}{l}
\xi \left( s,t\right) =A_{s}e^{p_{s}t}+B_{s}e^{p_{s}^{\prime }t} \\ 
\eta \left( s,t\right) =C_{s}e^{p_{s}t}+D_{s}e^{p_{s}^{\prime }t}%
\end{array}%
\right.
\end{equation*}

\noindent If $p_{s}$ and $p_{s}^{\prime }$ are real, then $A_{N-s}=\overline{%
A_{s}}$, etc. If $p_{s}$ and $p_{s}^{\prime }$ are conjugate complex
numbers, then $B_{N-s}=\overline{A_{s}}$, etc. It is straightforward to
verify that the coefficients satisfy the relations:

\begin{equation*}
\left\{ 
\begin{array}{l}
A_{s}\left( p_{s}-a+4\mu \sin ^{2}\left( \frac{\pi s}{N}\right) \right)
=bC_{s} \\ 
B_{s}\left( p_{s}^{\prime }-a+4\mu \sin ^{2}\left( \frac{\pi s}{N}\right)
\right) =bD_{s}%
\end{array}%
\right.
\end{equation*}

Now if $\limfunc{Max}\left( \limfunc{Re}\left( p_{s}\right) \right) >0$,
some diverging Fourier modes will become dominant and push the system out of
equilibrium. Such a possibility can happen only under precise conditions
relating the parameters $a,b,c,d$ of the internal chemical equilibrium to
the parameters $\mu ,\nu $ of the external spatial diffusion. Turing
explains very well that \emph{generically} only a single Fourier mode (with
its conjugate) can become dominant.\ Indeed, if it was not the case,

\begin{quotation}
\noindent \textquotedblleft the quantities $a,b,c,d,\mu ,\nu $ will be
restricted to satisfy some special condition, which they would be unlikely
to satisfy by chance.\textquotedblright ~(p.~50)
\end{quotation}

\noindent Let $s_{0}$ be the index yielding $\limfunc{Max}\left( \limfunc{Re}%
\left( p_{s}\right) \right) $ and suppose $\limfunc{Re}\left(
p_{s_{0}}\right) >0$. If the two eigenvalues $p_{s_{0}}$ and $%
p_{s_{0}}^{\prime }$ are real, the pair $\left( p_{s_{0}},p_{N-s_{0}}\right) 
$ will induce divergences since $\sin ^{2}\left( \frac{\pi \left(
N-s_{0}\right) }{N}\right) =\sin ^{2}\left( \frac{\pi s_{0}}{N}\right) $,
and if they are complex conjugate the two pairs $\left(
p_{s_{0}},p_{N-s_{0}}\right) $ and $\left( p_{s_{0}}^{\prime
},p_{N-s_{0}}^{\prime }\right) $ will both induce divergences.

\section{A toy model}

Turing presents a simple numerical example~p.~52. The parameters are

\begin{eqnarray*}
a &=&I-2,~b=2.5,~c=-1.25,~d=I+1.5 \\
\mu ^{\prime } &=&1,~\nu ^{\prime }=\frac{1}{2},~\frac{\mu }{\mu ^{\prime }}=%
\frac{\nu }{\nu ^{\prime }}=\left( \frac{N}{2\pi \rho }\right)
^{2},~U=\left( \frac{N}{\pi \rho }\right) ^{2}\sin ^{2}\left( \frac{\pi s}{N}%
\right)
\end{eqnarray*}

\noindent The characteristic equation is therefore


\begin{eqnarray*}
\left( p-a+4\mu \sin ^{2}\left( \frac{\pi s}{N}\right) \right) \left(
p-d+4\nu \sin ^{2}\left( \frac{\pi s}{N}\right) \right) &=&bc \\
\left( p-I+2+4\left( \frac{N}{2\pi \rho }\right) ^{2}\sin ^{2}\left( \frac{%
\pi s}{N}\right) \right) \left( p-I-1.5+2\left( \frac{N}{2\pi \rho }\right)
^{2}\sin ^{2}\left( \frac{\pi s}{N}\right) \right) &=&bc\\
\left( p-I+2+U\right) \left( p-I-1.5+\frac{1}{2}U\right) +\left( 2.5\right)
\left( 1.25\right) &=&0 \\
\left( p-I\right) ^{2}+\left( \frac{1}{2}+\frac{3}{2}U\right) \left(
p-I\right) +\frac{1}{2}\left( U-\frac{1}{2}\right) ^{2} &=&0
\end{eqnarray*}

\noindent We observe that $p=I$ for $U=\frac{1}{2}$. Let $s_{c}$ be the
corresponding value of $s$. If the radius $\rho $ of the ring is such that
there exists an integer $s_{0}$ satisfying $U=\left( \frac{N}{\pi \rho }%
\right) ^{2}\sin ^{2}\left( \frac{\pi s_{0}}{N}\right) =\frac{1}{2}$, then
there will exist stationary waves with $s_{0}$ lobes.\ Otherwise, it will be
the $s_{0}$ nearest to $s_{c}$ which will dominate.

\begin{figure}[tbp]
\begin{center}
\includegraphics[
width= 8.1012cm,height= 7.921cm]{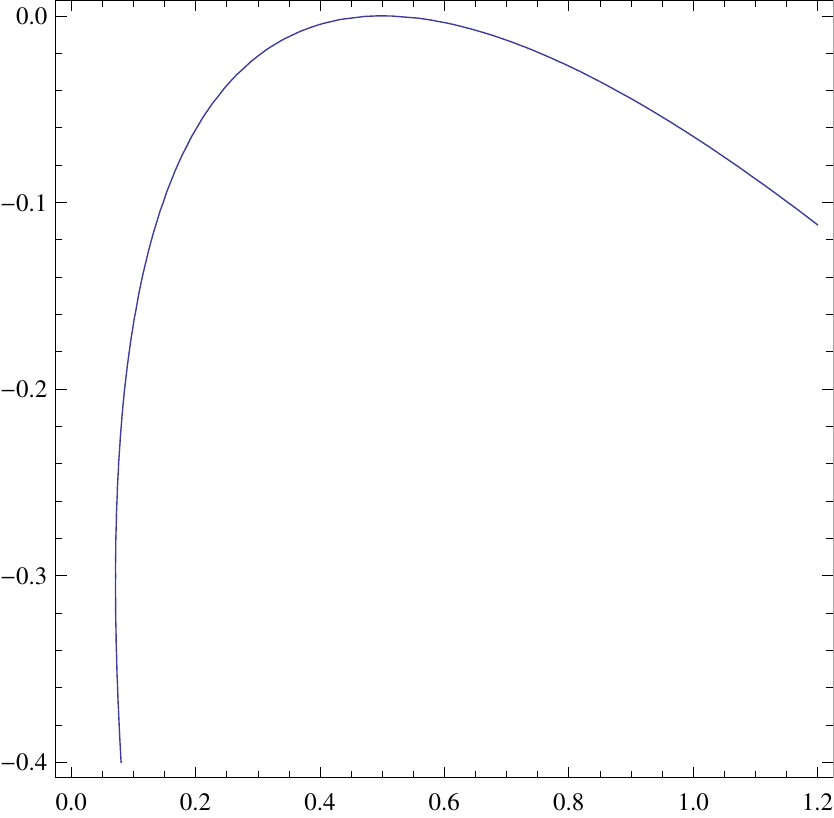}
\end{center}
\caption{The graph $\Gamma $ of $p^{2}+\left( \frac{1}{2}+\frac{3}{2}%
U\right) p+\frac{1}{2}\left( U-\frac{1}{2}\right) ^{2}=0$ for $U\in \left[
0,1.2\right] $ and $p\in \left[ -0.4,0\right] $. }
\label{Turing_fig1_Mathematica}
\end{figure}

The figure \ref{Turing_fig1_Mathematica} displays for $I=0$ the graph $%
\Gamma $ of the hyperbola $p^{2}+\left( \frac{1}{2}+\frac{3}{2}U\right) p+%
\frac{1}{2}\left( U-\frac{1}{2}\right) ^{2}=0$ in the $\left( U,p\right) $
plane for $U\in \left[ 0,1.2\right] $ and $p\in \left[ -0.4,0\right] $, and $%
I=0$.



The points of $\Gamma $ are evident.\ If $p$ is considered as a parameter, 
\begin{equation*}
U=\frac{1}{2}\left( 1-3p\pm \sqrt{p^{2}-10p}\right)
\end{equation*}

\noindent and if $U$ is considered as a parameter, 
\begin{equation*}
p=\frac{1}{4}\left( -1-3U\pm \sqrt{U^{2}+14U-1}\right) .
\end{equation*}

\noindent The solutions of $U^{2}+14U-1=0$ are $U=-7\pm 5\sqrt{2}$ but, as $%
U $ is a real square $\left( \frac{N}{\pi \rho }\right) ^{2}\sin ^{2}\left( 
\frac{\pi s}{N}\right) $, the only admissible value is $U_{c}=-7+5\sqrt{2}%
\sim 0.071$ and for $U_{c}$ the two values of $p$ are equal to $-\frac{1}{4}%
\left( 1+3U\right) \sim 0.30325$. For $U>U_{c}$, the two $p$ roots are real
and for $0\leq U<U_{c}$, they have an imaginary part $\func{Im}\left(
p\right) =\pm \sqrt{U^{2}+14U-1}$ while the real part move on the segment $%
\limfunc{Re}\left( p\right) =-\frac{1}{4}\left( 1+3U\right) $ from the point 
$\left( U=0,p=-\frac{1}{4}\right) $ to the point $\left( U\sim 0.071,p\sim
0.30325\right) $. In what concerns $p$, as $U$ is real, we must have $\sqrt{%
p^{2}-10p}$ real, that is $p^{2}-10p\geq 0$ i.e. $p\in \left( -\infty ,0%
\right] $ or $p\in \left[ 10,+\infty \right) $.

\begin{figure}[ptb]
\begin{center}
\includegraphics[
width= 7.0973cm,height= 6.2802cm]{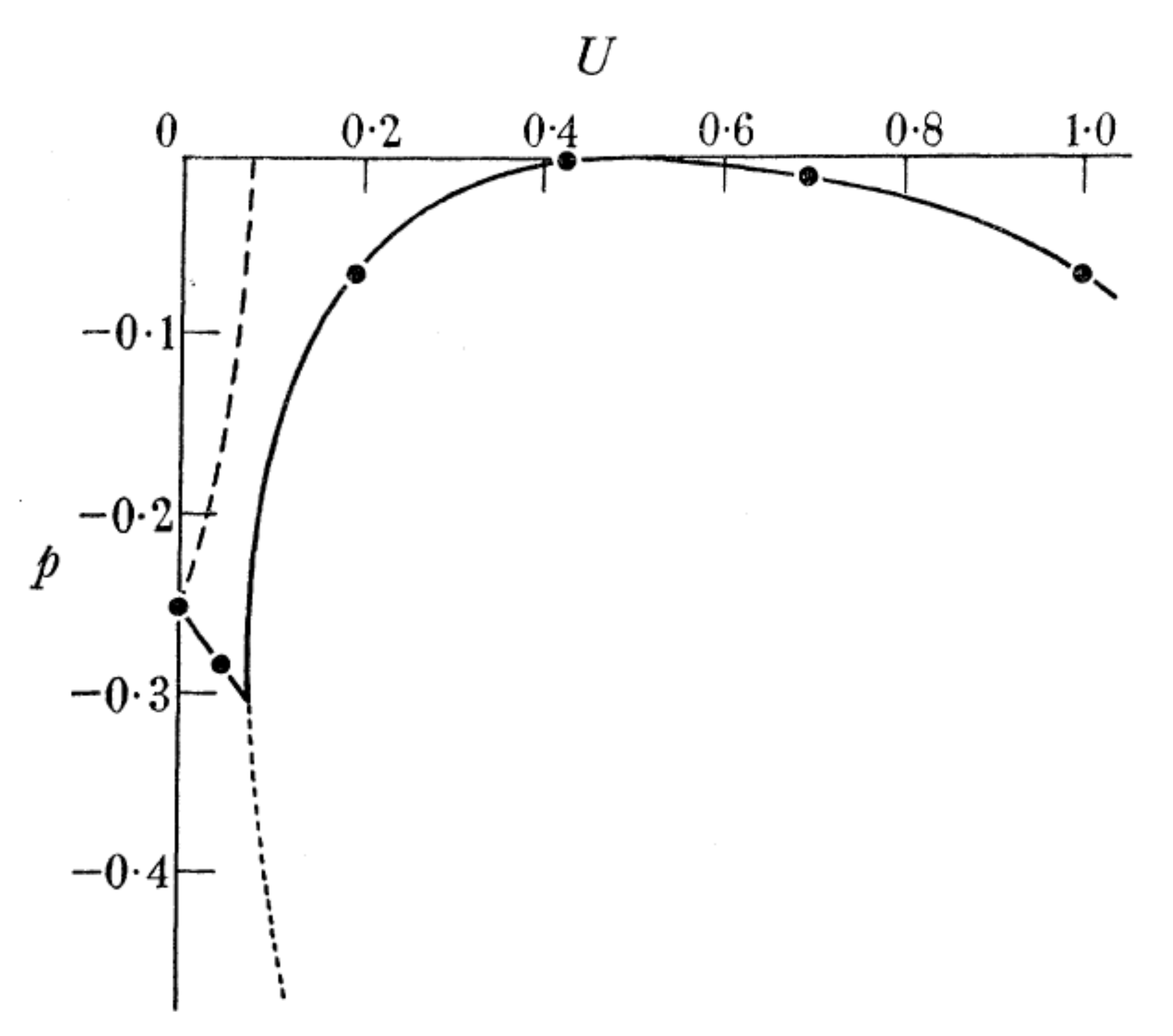}
\end{center}
\caption{Turing's figure 1 which displays $\limfunc{Re}\left( p\right) $ and 
$-\left\vert \func{Im}\left( p\right) \right\vert $ as functions of $U$ for $%
I=0$. The full line and the dotted line represent respectively $\limfunc{Re}%
\left( p_{s}\right) $ and $\limfunc{Re}\left( p_{s}^{\prime }\right) $,
while the broken line represents $-\left\vert \func{Im}\left( p\right)
\right\vert $.\ Turing has indicated with black thick points the integer
values from $s=0$ (left) to $s=5$ (right). }
\label{Turing_fig1}
\end{figure}

The figure \ref{Turing_fig1} reproduces Turing's figure 1 which displays $%
\limfunc{Re}\left( p\right) $ and $-\left\vert \func{Im}\left( p\right)
\right\vert $ as functions of $U$ for $I=0$.



\section{Conditions for instability and the critical point}

In the \S 9 \textquotedblleft Further considerations on the mathematics of
the ring\textquotedblright , Turing analyzes further the conditions under
which a diffusion-driven instability can occur. We will present and complete
his computations using the \emph{continuous} model which is easier to
understand.

As reaction-diffusion equations are linear, and since every function on $%
S^{1}$ is a linear superposition of harmonics $e^{ik\theta }$, we look at
solutions of the form

\begin{equation*}
\left( 
\begin{array}{c}
x \\ 
y%
\end{array}%
\right) =e^{\lambda t}e^{ik\theta }\left( 
\begin{array}{c}
x_{0} \\ 
y_{0}%
\end{array}%
\right)
\end{equation*}

\noindent which implies immediately

\begin{gather*}
\left( 
\begin{array}{c}
\dot{x} \\ 
\dot{y}%
\end{array}%
\right) =\lambda e^{\lambda t}e^{ik\theta }\left( 
\begin{array}{c}
x_{0} \\ 
y_{0}%
\end{array}%
\right) = \\
J_{0}e^{\lambda t}e^{ik\theta }\left( 
\begin{array}{c}
x_{0} \\ 
y_{0}%
\end{array}%
\right) +\left( 
\begin{array}{cc}
\mu ^{\prime } & 0 \\ 
0 & \nu ^{\prime }%
\end{array}%
\right) \left( -k^{2}e^{\lambda t}e^{ik\theta }\right) \left( 
\begin{array}{c}
x_{0} \\ 
y_{0}%
\end{array}%
\right) ,\text{ that is } \\
\left( J_{0}-k^{2}D-\lambda I\right) \left( 
\begin{array}{c}
x \\ 
y%
\end{array}%
\right) =0,\text{ with }D=\left( 
\begin{array}{cc}
\mu ^{\prime } & 0 \\ 
0 & \nu ^{\prime }%
\end{array}%
\right)
\end{gather*}

\noindent The characteristic equation is therefore

\begin{equation*}
\func{Det}\left( J_{0}-k^{2}D-\lambda I\right) =\func{Det}\left( 
\begin{array}{cc}
a-\mu ^{\prime }k^{2}-\lambda & b \\ 
c & d-\nu ^{\prime }k^{2}-\lambda%
\end{array}%
\right) =0
\end{equation*}

\noindent and the dispersion relations are

\begin{equation*}
\left( \lambda -a+\mu ^{\prime }k^{2}\right) \left( \lambda -d+\nu ^{\prime
}k^{2}\right) =bc
\end{equation*}

\noindent or, if we write this equation $\lambda ^{2}-S\left( k^{2}\right)
\lambda +P\left( k^{2}\right) =0$ with $S\left( k^{2}\right) $ the sum of
its two roots and $P\left( k^{2}\right) $ their product,

\begin{equation*}
\lambda ^{2}-\left( \limfunc{Tr}\left( J_{0}\right) -k^{2}\limfunc{Tr}\left(
D\right) \right) \lambda +\left( \mu ^{\prime }\nu ^{\prime }k^{4}-\left(
\nu ^{\prime }a+\mu ^{\prime }d\right) k^{2}+\func{Det}\left( J_{0}\right)
\right) =0
\end{equation*}

We want $\limfunc{Tr}\left( J_{0}\right) =a+d<0$ and $\func{Det}\left(
J_{0}\right) =ad-bc>0$ to ensure the stability of the internal chemical
equilibrium. But we want also one $\lambda $ with $\limfunc{Re}\left(
\lambda \right) >0$ to ensure a diffusion-driven instability. As $\limfunc{Tr%
}\left( J_{0}\right) <0$ and $\limfunc{Tr}\left( D\right) =\mu ^{\prime
}+\nu ^{\prime }>0$, this implies $S\left( k^{2}\right) =\limfunc{Tr}\left(
J_{0}\right) -k^{2}\limfunc{Tr}\left( D\right) <0$. If $P\left( k^{2}\right) 
$ happened to be $>0$, we would have two roots with $\limfunc{Re}\left(
\lambda \right) <0$, so we must have $P\left( k^{2}\right) <0$. But as $%
\func{Det}\left( J_{0}\right) >0$ by hypothesis and of course $\mu ^{\prime
}\nu ^{\prime }k^{4}>0$, we need in fact $\nu ^{\prime }a+\mu ^{\prime
}d>\mu ^{\prime }\nu ^{\prime }k^{2}+\func{Det}\left( J_{0}\right) /k^{2}>0$%
. This is a first condition. As $\limfunc{Tr}\left( J_{0}\right) =a+d<0$ and 
$\mu ^{\prime },\nu ^{\prime }>0$, we need $\mu ^{\prime }\neq \nu ^{\prime
} $ since, if $\mu ^{\prime }$ and $\nu ^{\prime }$ would be equal, we would
have $\nu ^{\prime }a+\mu ^{\prime }d=\mu ^{\prime }\left( a+d\right) <0$.

It is essential to strongly emphasize here the fact that the diffusibility
of the two morphogens $X$ and $Y$ must be \emph{sufficiently different} in
order that an instability can occur. It is the key of Turing's discovery.

In the example of \S 10, $\mu ^{\prime }=\alpha \frac{1}{2}$, $\nu ^{\prime
}=\alpha \frac{1}{4}$ $\left( \alpha >0\right) $, $a=-2$, $d=\frac{3}{2}$
and the condition $\nu ^{\prime }a+\mu ^{\prime }d=\alpha \left( -2\times 
\frac{1}{4}+\frac{3}{2}\times \frac{1}{2}\right) =\frac{\alpha }{4}>0$ is
therefore satisfied.

There is a second condition.\ $P\left( k^{2}\right) =\mu ^{\prime }\nu
^{\prime }k^{4}-\left( \nu ^{\prime }a+\mu ^{\prime }d\right) k^{2}+\func{Det%
}\left( J_{0}\right) $ is a second degree polynomial in $k^{2}$ and we want
it to become $<0$ for values of $k^{2}$ which must necessarilly be positive
since $k^{2}$ is a real square. Let $\delta =\frac{\mu ^{\prime }}{\nu
^{\prime }}$. The graph of $P\left( k^{2}\right) $ is a parabola $\Pi $
starting at $\func{Det}\left( J_{0}\right) >0$ for $k=0$. If $\delta <\delta
_{c}$ for a critical value to be computed, $\Pi $ is over the $k^{2}$-axis
and the condition $P\left( k^{2}\right) <0$ cannot be satisfied. But if $%
\delta >\delta _{c}$, $\Pi $ intersects the $k^{2}$-axis at two points $%
k_{1}^{2}$ and $k_{2}^{2}>k_{1}^{2}$ and inside the interval $\left[
k_{1}^{2},k_{2}^{2}\right] $ the condition $P\left( k^{2}\right) <0$ is
satisfied: there exists an eigenvalue $\lambda $ with $\limfunc{Re}\left(
\lambda \right) >0$.

The computation of $\delta _{c}$ is rather tedious. Let $u=k^{2}$.\ The
polynomial $P\left( u\right) $ and its first and second derivatives are

\begin{eqnarray*}
P\left( u\right) &=&\mu ^{\prime }\nu ^{\prime }u^{2}-\left( \nu ^{\prime
}a+\mu ^{\prime }d\right) u+ad-bc \\
P^{\prime }\left( u\right) &=&-\left( \nu ^{\prime }a+\mu ^{\prime }d\right)
+2\mu ^{\prime }\nu ^{\prime }u \\
P^{\prime \prime }\left( u\right) &=&2\mu ^{\prime }\nu ^{\prime }
\end{eqnarray*}

\noindent So the minimum of $\Pi $ is given by $P^{\prime }\left( u\right)
=-\left( \nu ^{\prime }a+\mu ^{\prime }d\right) +2\mu ^{\prime }\nu ^{\prime
}u=0$ and it is a minimum since $P^{\prime \prime }\left( u\right) =2\mu
^{\prime }\nu ^{\prime }>0$. Its value is

\begin{equation*}
u_{0}=k_{0}^{2}=\frac{\nu ^{\prime }a+\mu ^{\prime }d}{2\mu ^{\prime }\nu
^{\prime }}
\end{equation*}

\noindent and we have

\begin{equation*}
P\left( k_{0}^{2}\right) =ad-bc-\frac{1}{4}\frac{\left( \nu ^{\prime }a+\mu
^{\prime }d\right) ^{2}}{\mu ^{\prime }\nu ^{\prime }}
\end{equation*}

\noindent For $P\left( k^{2}\right) $ to become $<0$, we must have therefore

\begin{equation*}
0<ad-bc<\frac{1}{4}\frac{\left( \nu ^{\prime }a+\mu ^{\prime }d\right) ^{2}}{%
\mu ^{\prime }\nu ^{\prime }}=\frac{1}{4}\frac{\left( a+\delta d\right) ^{2}%
}{\delta }
\end{equation*}

\noindent and $\delta _{c}$ is given by the equation

\begin{equation*}
ad-bc=\frac{1}{4}\frac{\left( a+\delta _{c}d\right) ^{2}}{\delta _{c}}
\end{equation*}

In the example, when $\gamma =0$, we have $ad-bc=\frac{1}{8}$, $\nu ^{\prime
}a+\mu ^{\prime }d=\frac{\alpha }{4}$, $\mu ^{\prime }\nu ^{\prime }=\frac{%
\alpha ^{2}}{8}$, and we verify that $\frac{1}{8}=\frac{1}{4}\times \left( 
\frac{1}{4}\right) ^{2}\times 8$.\ So Turing's system is at its \emph{%
critical point} for $\gamma =0$.

Let us now investigate more precisely the second degree equation giving $%
\delta _{c}$.\ It can be written

\begin{equation*}
d^{2}\delta _{c}^{2}+2\left( 2bc-ad\right) \delta _{c}+a^{2}=0
\end{equation*}

\noindent and its two solutions are

\begin{equation*}
\delta _{c\pm }=\frac{ad-2bc\pm 2\sqrt{-bc\left( ad-bc\right) }}{d^{2}}
\end{equation*}

\noindent But as one root, and hence both roots, must be real, we need $%
-bc\left( ad-bc\right) >0$ and, as $ad-bc>0$ by hypothesis, we must have $%
bc<0$.

In the example, for $\gamma =0$, we have effectively $-\frac{25}{16}\times
2<0$ and

\begin{equation*}
\delta _{c}=\left( \frac{1}{32}\right) ^{2}\left[ \frac{1}{8}-\left( -\frac{%
25}{16}\times 2\right) +2\sqrt{\left( \frac{25}{16}\right) \times \left( 
\frac{1}{8}\right) }\right] =2=\delta
\end{equation*}

\noindent As $\delta =\delta _{c}$ we are indeed at the critical point. Then

\begin{equation*}
P\left( k^{2}\right) =\frac{\alpha ^{2}}{8}k^{4}-\frac{\alpha }{4}k^{2}+%
\frac{1}{8}=\frac{1}{8}\left( \alpha k^{2}-1\right) ^{2}
\end{equation*}

\noindent and the parabola $\Pi $ is tangent at the $k^{2}$-axis and at the
point of tangency the eigenvalues are $\lambda _{+}=0$ and $\lambda _{\_}=%
\limfunc{Tr}\left( J_{0}\right) -k^{2}\limfunc{Tr}\left( D\right) =-\left( 
\frac{3\alpha }{4}k^{2}\right) -\frac{1}{2}<0$.\ So, at the crossing of the
critical point, the eigenvalue $\lambda _{+}$ becomes $>0$.

\section{The bifurcation}

After having analyzed the conditions of a diffusion-driven instability at a
critical point, Turing analyzed further the behavior of the system in the
neighbourhood of the critical point.\ To this end, he varied the small
parameter $\gamma $ around its critical value $\gamma =0$. Today,
computations are very easy, but at his time they were difficult and he must
use the computer he had himself constructed. We will first do them for the
continuous model and then return to Turing's own discrete model.

\subsection{Continuous model}

The chemical internal equilibrium is now given by the concentrations of
morphogens

\begin{equation*}
X_{0}=\frac{1}{7}\left( -25+\sqrt{2^{10}+7\times 55\gamma }\right) ,~Y_{0}=1
\end{equation*}

\noindent We linearize the system in the neighbourhood of $\left(
X_{0},Y_{0}\right) $ and compute first order expansions in the small
parameter $\varepsilon =\frac{7\times 55}{2^{10}}\gamma =\frac{385}{1024}%
\gamma $. We get

\begin{equation*}
a=-2-\varepsilon ,b=-\frac{25}{16}-\frac{25}{7}\varepsilon ,c=2+\varepsilon
,d=\frac{3}{2}+\frac{25}{7}\varepsilon
\end{equation*}

\noindent and the conditions for instability: $\limfunc{Tr}\left(
J_{0}\right) =a+d=-\frac{1}{2}+\frac{18}{7}\varepsilon $ must be $<0$, which
implies $\varepsilon <\frac{7}{36}\sim 0.194$; $\nu ^{\prime }a+\mu ^{\prime
}d=\frac{1}{4}+\frac{43}{28}\varepsilon $ must be $>0$, which implies $%
\varepsilon >-\frac{7}{43}$; $\func{Det}\left( J_{0}\right) =ad-bc=\frac{1}{8%
}+\frac{1}{16}\varepsilon $ must be $>0$, which implies $\varepsilon >-2$; $%
ad$ must be $<0$, which implies $\varepsilon >-\frac{42}{121}$; $ad-bc$ must
be $<\frac{1}{4}\frac{\left( \nu ^{\prime }a+\mu ^{\prime }d\right) ^{2}}{%
\mu ^{\prime }\nu ^{\prime }}$, which implies $\varepsilon >0$.

The equation yielding the eigenvalues is now

\begin{equation*}
\lambda ^{2}-\left( \limfunc{Tr}\left( J_{0}\right) -k^{2}\limfunc{Tr}\left(
D\right) \right) \lambda +P\left( k^{2}\right) =0
\end{equation*}

\begin{figure}[ptb]
\begin{center}
\includegraphics[
width= 8.1012cm,height= 4.8194cm]{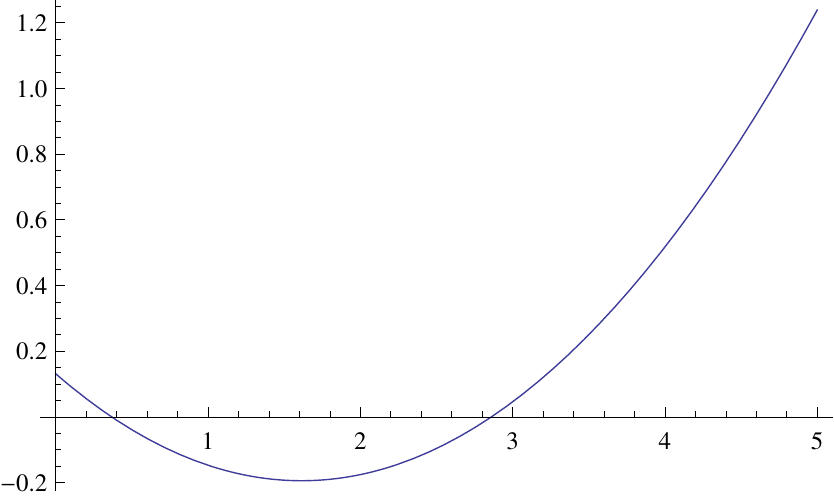}
\end{center}
\caption{The graph of $P\left( k^{2}\right) $ for $\protect\varepsilon =0.1$%
. Inside the interval $\left[ 0.367,2.862\right] $ of $k^{2} $ we have $%
P\left( k^{2}\right) <0$.}
\label{P_k2_e_01}
\end{figure}

Figure \ref{P_k2_e_01} displays the graph of $P\left( k^{2}\right) $ for $%
\varepsilon =0.1.\ $The roots of $P\left( k^{2}\right) =0$ are $0.367$ and $%
2.862$ and inside their interval we have $P\left( k^{2}\right) <0$.



The characteristic equation is

\begin{equation*}
\lambda ^{2}-\left( \limfunc{Tr}\left( J_{0}\right) -k^{2}\limfunc{Tr}\left(
D\right) \right) \lambda +\left( \mu ^{\prime }\nu ^{\prime }k^{4}-\left(
\nu ^{\prime }a+\mu ^{\prime }d\right) k^{2}+\func{Det}\left( J_{0}\right)
\right) =0
\end{equation*}

\noindent that is

\begin{equation*}
\lambda ^{2}+\left( \frac{1}{2}+\frac{3k^{2}}{4}-\frac{18}{7}\varepsilon
\right) \lambda +\left( \frac{k^{4}}{8}-\left( \frac{1}{4}+\frac{43}{28}%
\varepsilon \right) k^{2}+\frac{1}{8}+\frac{1}{16}\varepsilon \right) =0
\end{equation*}

\noindent and its roots $\lambda _{\pm }$ are approximated by 
\begin{equation*}
\frac{1}{128}\left( -7+36\varepsilon \pm \sqrt{-49-553\varepsilon
+1296\varepsilon ^{2}+490k^{2}-308\varepsilon k^{2}+343k^{4}}\right)
\end{equation*}

\begin{figure}[ptb]
\begin{center}
\includegraphics[
width= 8.1012cm,height= 4.9468cm]{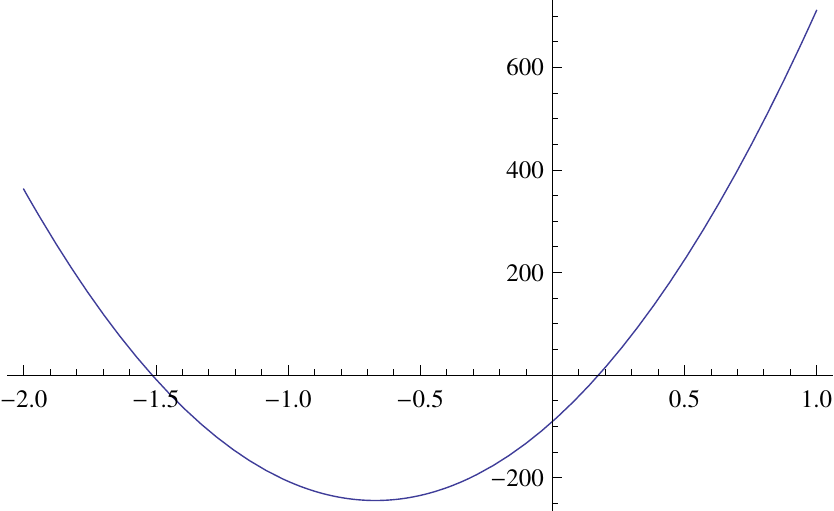}
\end{center}
\caption{The discriminant $\Delta \left( k^{2}\right) $ for $\protect%
\varepsilon =0.1$.\ $\Delta $ is negative inside the open interval $\left]
-1.5146,0.1758\right[ .$}
\label{Delta_e_01}
\end{figure}

\begin{figure}[ptb]
\begin{center}
\includegraphics[
width= 8.1012cm,height= 5.0478cm]{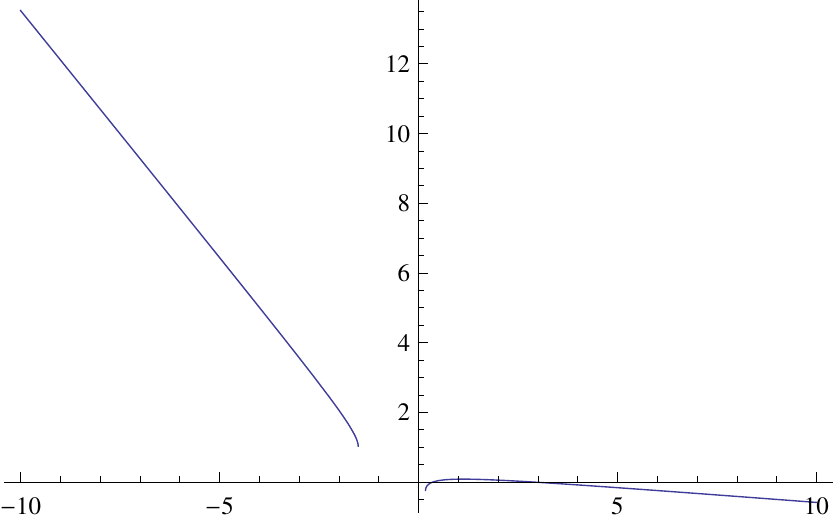}
\end{center}
\caption{Graph of $\protect\lambda _{+}$ (including the irrelevant negative $%
k^{2}$-axis). $\protect\lambda _{+}\geq 0$ for $k^{2}\in \left[ 0.367,2.862%
\right] $. There is no graph inside the open interval $\left] -1.5146,0.1758%
\right[ $ where the discriminant $\Delta $ of the characteristic equation is 
$<0$.}
\label{Lam2}
\end{figure}

\begin{figure}[ptb]
\begin{center}
\includegraphics[
width= 8.1012cm,height= 4.8348cm]{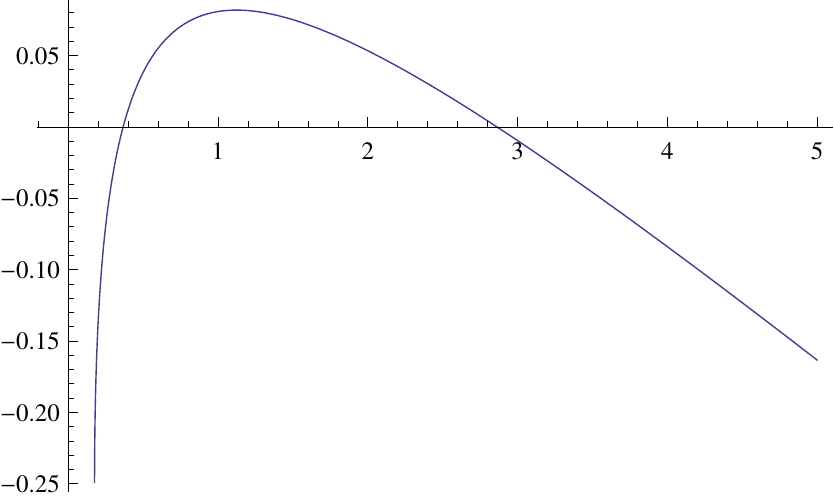}
\end{center}
\caption{Zoom on the interval $k^{2}\in \left[ 0.367,2.862\right] $ of the
figure \protect\ref{Lam2} where $\protect\lambda _{+}\geq 0$. }
\label{Lam2_zoom}
\end{figure}

\begin{figure}[tbp]
\begin{center}
\includegraphics[
width= 8.1012cm,height= 4.9644cm]{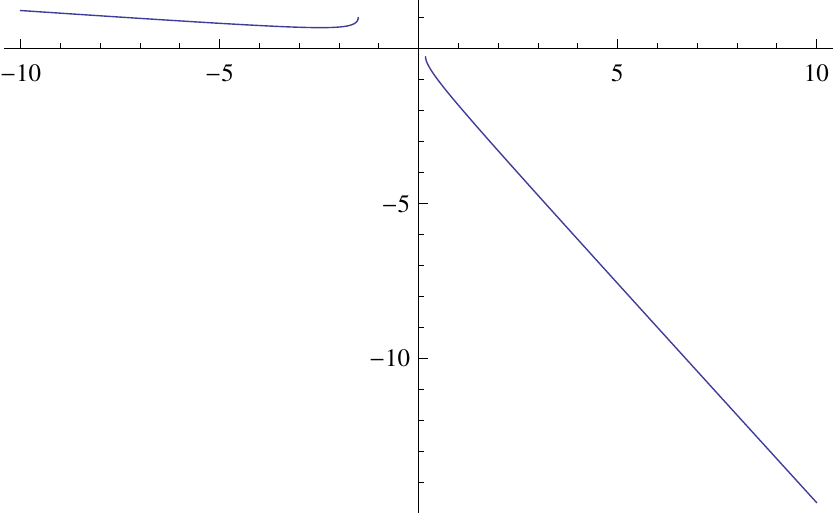}
\end{center}
\caption{Graph of $\lambda _{\_}$ (including the irrelevant
negative $k^{2}$-axis). $\lambda _{\_}\ $is always $<0$ for $k^{2}>0$. There is no graph
inside the open interval $\left] -1.5146,0.1758\right[ $ where the
discriminant $\Delta $ of the characteristic equation is $<0$.}
\label{Lam1}
\end{figure}

\noindent Figures \ref{Lam2} and \ref{Lam1} display the graphs of $\lambda
_{+}$ and $\lambda _{\_}$ (including the irrelevant negative $k^{2}$-axis).
For $\lambda _{+}$ we see that $\lambda _{+}\geq 0$ for $k^{2}\in \left[
0.367,2.862\right] $. Figure \ref{Lam2_zoom} zooms on this interval.\ There
is no graph inside the open interval $\left] -1.5146,0.1758\right[ $ where
the discriminant $\Delta $ of the characteristic equation is $<0$ (see
figure \ref{Delta_e_01}).



\subsection{Discrete model}

Let us come back to the discrete ring model composed of $N=20$ cells. At the
critical point $\gamma =0$, the $20$ characteristic equations in the Fourier
domain are

\begin{equation*}
\left( p+2+2\sin ^{2}\left( \frac{\pi s}{20}\right) \right) \left(
p-1.5+\sin ^{2}\left( \frac{\pi s}{20}\right) \right) +\frac{25}{8}=0
\end{equation*}

\begin{figure}[ptb]
\begin{center}
\includegraphics[
width= 14.1243cm,height= 4.5426cm]{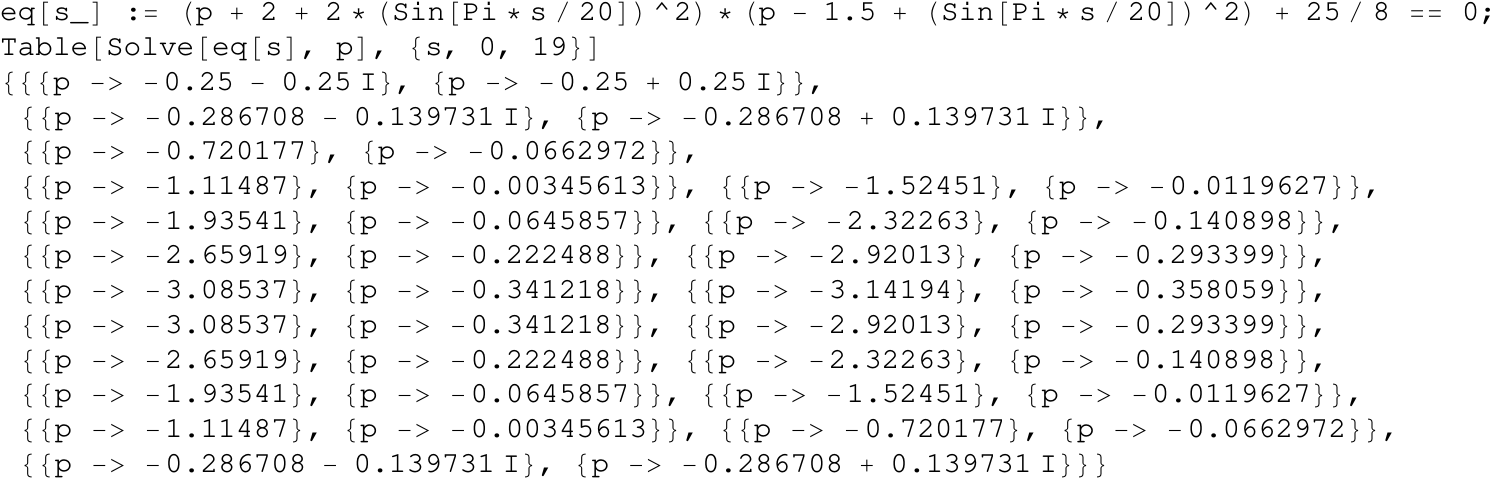}
\end{center}
\caption{The table of the $20$ pairs $\left( p_{s},p_{s}^{\prime }\right) $
of eigenvalues of the discrete ring model for $s=0,\ldots ,19$ and $\protect%
\gamma =0$. }
\label{EigenVal_Ring_0}
\end{figure}

\noindent The figure \ref{EigenVal_Ring_0} shows the table of the $20$ pairs 
$\left( p_{s},p_{s}^{\prime }\right) $ of eigenvalues for $s=0,\ldots ,19$.
We see that if we order the $p$ w.r.t to increasing $\limfunc{Re}\left(
p\right) $ we get $p_{3}=p_{17}=-0.00346$, $p_{4}=p_{16}=-0.012$, $%
p_{5}=p_{15}=-0.064$, $p_{2}=p_{18}=-0.066$.\ It is therefore the mode $%
p_{3} $ which can most readily become $>0.$



Let us now vary the \emph{small slow parameter} $\gamma .$ Turing varied $%
\gamma $ almost adiabatically from $-\frac{1}{4}$ (stability) to $\frac{1}{16%
}$ (instability) at speed $\dot{\gamma}=2^{-7}=\frac{1}{128}$. This
corresponds to variations $\varepsilon :-0.094\rightarrow 0.0235$ for $%
\varepsilon $ and $t:0\rightarrow 40$ for the discrete time $t$. For $\gamma
=-\frac{1}{4}$, the equilibrium is $\left( X_{0}=0.78,Y_{0}=1\right) $, and $%
a_{0}=-1.9$, $b_{0}=-1.218$, $c_{0}=1.9$, $d_{0}=1.156$, and all $\limfunc{Re%
}\left( p_{s}\right) <0$: the system is stable.\ On the contrary, for $%
\gamma =-\frac{1}{16}$, the equilibrium is $\left(
X_{1}=1.053,Y_{1}=1\right) $, and $a_{1}=-2.023$, $b_{1}=-1.646$, $%
c_{1}=2.023$, $d_{1}=1.583$, and $p_{3}=p_{17}=0.0224>0$, $%
p_{4}=p_{16}=0.012>0$: the system is unstable. The figure \ref%
{EigenVal_Ring_g} shows the table of the $20$ pairs $\left(
p_{s},p_{s}^{\prime }\right) $ of eigenvalues for $s=0,\ldots ,19$ for $%
\gamma =\frac{1}{16}$.



\begin{figure}[ptb]
\begin{center}
\includegraphics[
width= 14.1221cm,height= 4.6854cm]{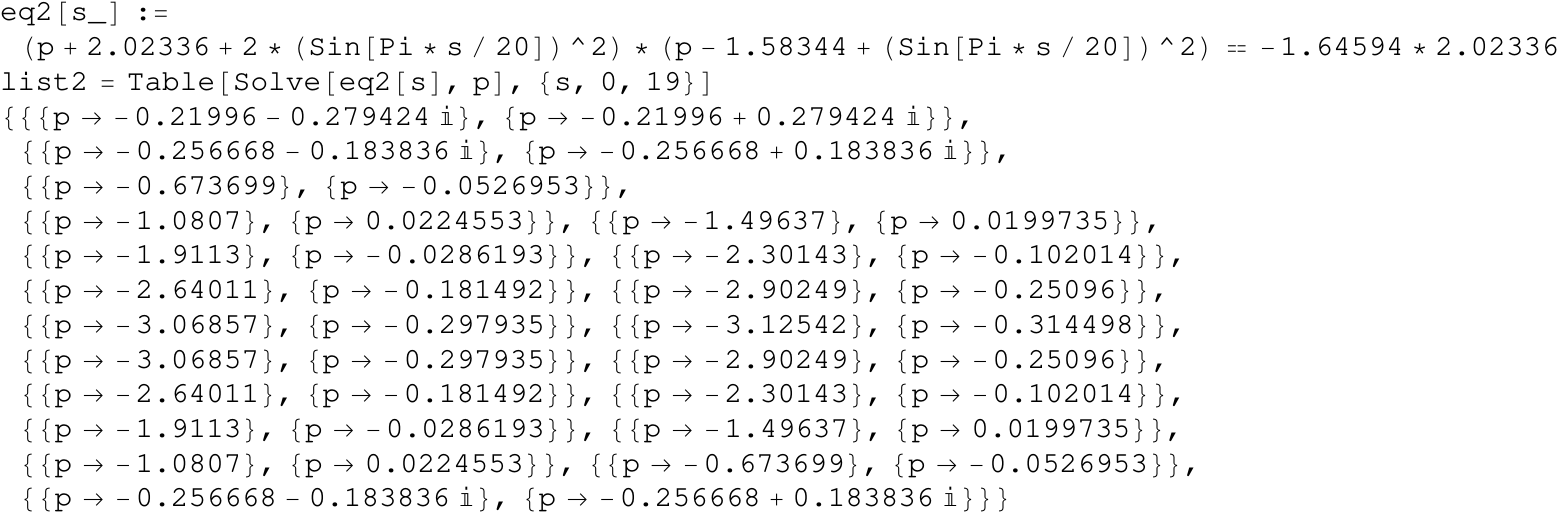}
\end{center}
\caption{The table of the $20$ pairs $\left( p_{s},p_{s}^{\prime }\right) $
of eigenvalues of the discrete ring model for $s=0,\ldots ,19$ and $\protect%
\gamma =\frac{1}{16}$.}
\label{EigenVal_Ring_g}
\end{figure}

\begin{figure}[ptb]
\begin{center}
\includegraphics[
width= 15.1281cm,height= 12.0748cm]{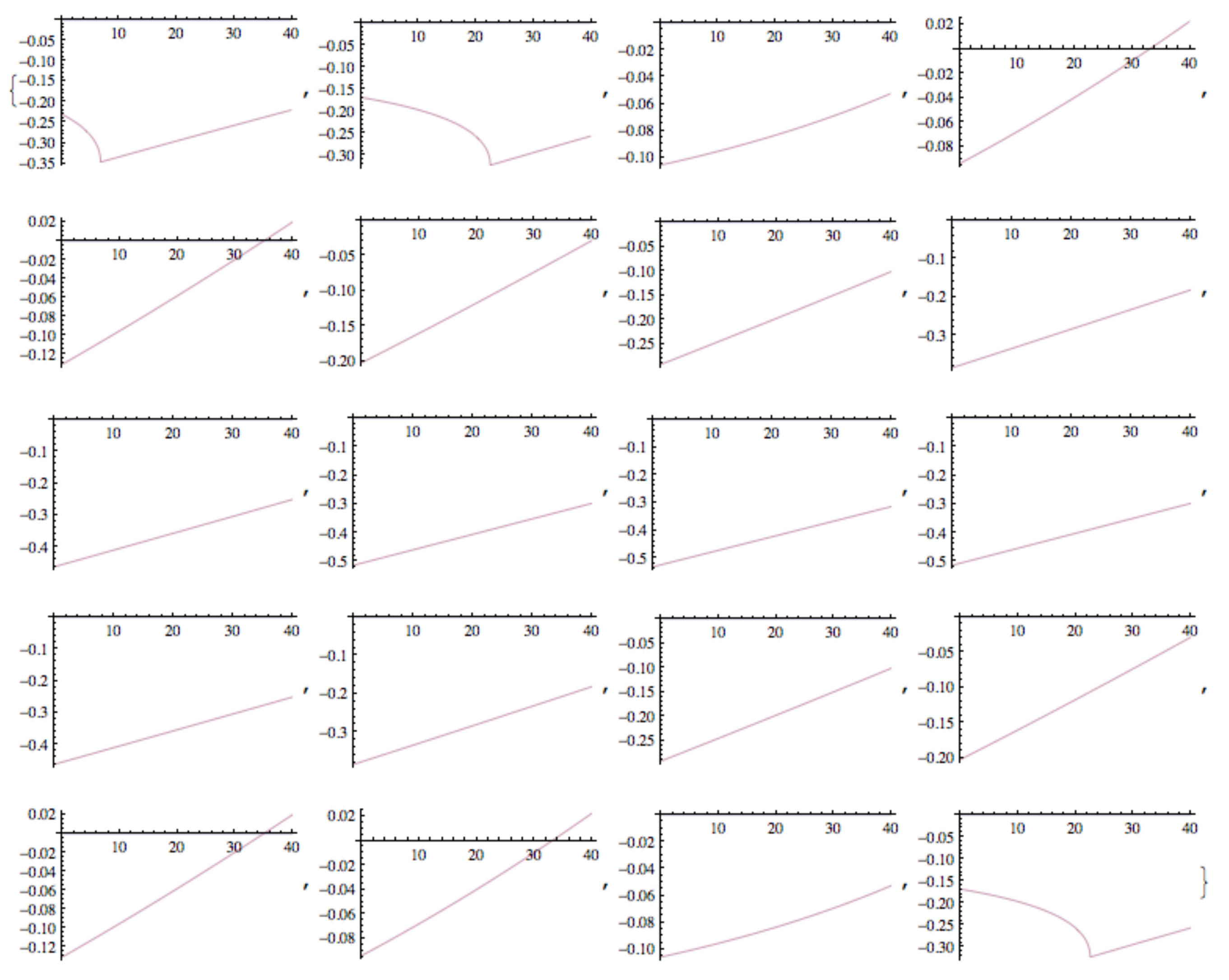}
\end{center}
\caption{The $20$ graphs $\limfunc{Re}\left( p_{s}\right) $ as functions of $%
t$ for $\protect\gamma $ varying from $-\frac{1}{4}$ to $\frac{1}{16}$. The
time $t$ varies from $t=0$ to $t=40$. The $\limfunc{Re}\left( p_{s}\right) $
become $>0$ for $s=3,17,4,16$.\ At $t=40$, $p_{3}=p_{17}=0.0224$ and $%
p_{4}=p_{16}=0.012$. For $s=0,1,19$, $\limfunc{Re}\left( p_{s}\right) $
presents an angular point because $p_{s}$ is a complex number with $\func{Im}%
\left( p_{s}\right) \neq 0$.\ It is the same phenomenon as in the toy model
of figure \protect\ref{Turing_fig1}.}
\label{Table_Re_p}
\end{figure}

In figure \ref{Table_Re_p} we show the $20$ graphs $\limfunc{Re}\left(
p_{s}\right) $ as functions of $t$ for $\gamma $ varying from $-\frac{1}{4}$
to $\frac{1}{16}$. The time $t$ varies from $t=0$ to $t=40$. We see the $%
\limfunc{Re}\left( p_{s}\right) $ which become $>0$ for $s=3,17,4,16$.\ At $%
t=40$, $p_{3}=p_{17}=0.0224$ and $p_{4}=p_{16}=0.012$. For $s=0,1,19$, $%
\limfunc{Re}\left( p_{s}\right) $ presents an angular point because $p_{s}$
is a complex number with $\func{Im}\left( p_{s}\right) \neq 0$.\ It is the
same phenomenon as in the toy model of figure \ref{Turing_fig1}. Figure \ref%
{Eigenval_Re_Im} shows the graphs of $\limfunc{Re}\left( p\right) $ and $%
\func{Im}\left( p\right) $ in such a case.



\begin{figure}[ptb]
\begin{center}
\includegraphics[
width= 8.1012cm,height= 5.0478cm]{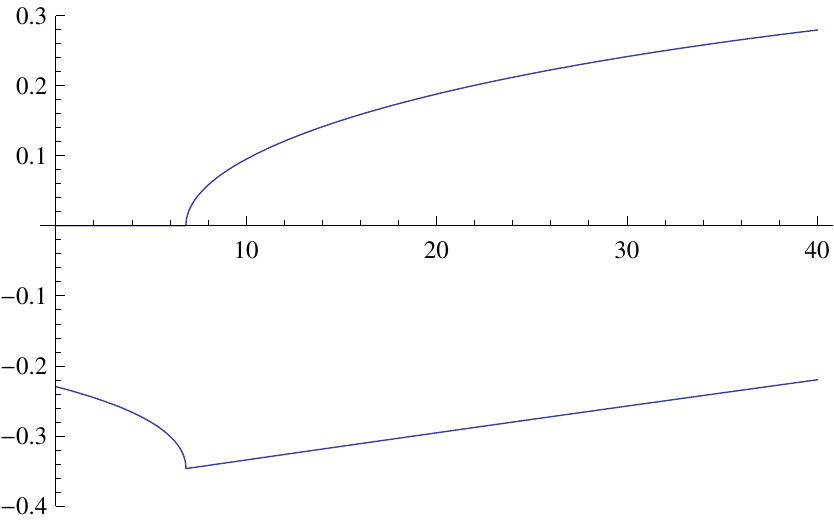}
\end{center}
\caption{When $\func{Im}\left( p\right) \neq 0$ (graph up), $\limfunc{Re}%
\left( p\right) $ (graph down) presents an angular point. }
\label{Eigenval_Re_Im}
\end{figure}

In his paper, Turing computes (with his recently constructed Manchester
computer) the table of the evolution of the ring (see figure \ref%
{Turing_Table1}) and shows how Fourier modes become dominant after the
bifurcation induced by the diffusion-driven instability (see figure \ref%
{Turing_fig3}). In the initial state, all cells are, up to small
fluctuations, in the equilibrium state $\left( X_{0}=1,Y_{0}=1\right) $.
After the bifurcation, a stationary oscillatory wave pattern with $3$ lobes
develops.\ The divergences induced by the instability are tamed by two
factors: (i) the concentration $X$ cannot become $<0$ and when $X\left(
r,t\right) $ vanishes, the process stops \emph{locally}, (ii) saturation
non-linear effects allow a new equilibrium to occur. These results
constitute a great achievement.



\begin{figure}[tbp]
\begin{center}
\includegraphics[
width= 15.1281cm,height= 9.0611cm]{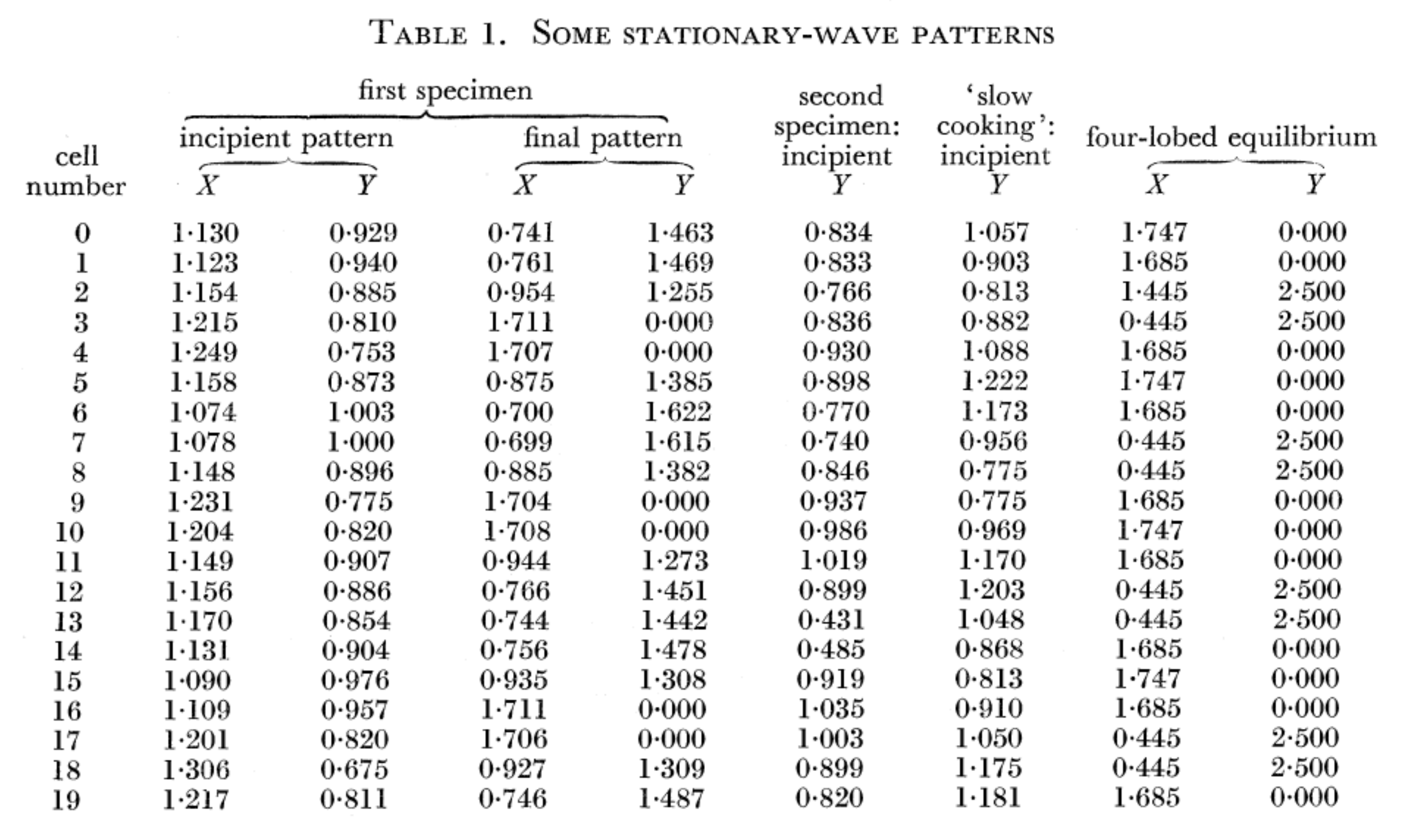}
\end{center}
\caption{Turing's computation of the evolution of the ring. }
\label{Turing_Table1}
\end{figure}



\begin{figure}[ptb]
\begin{center}
\includegraphics[
width= 12.1166cm,height= 10.0078cm]{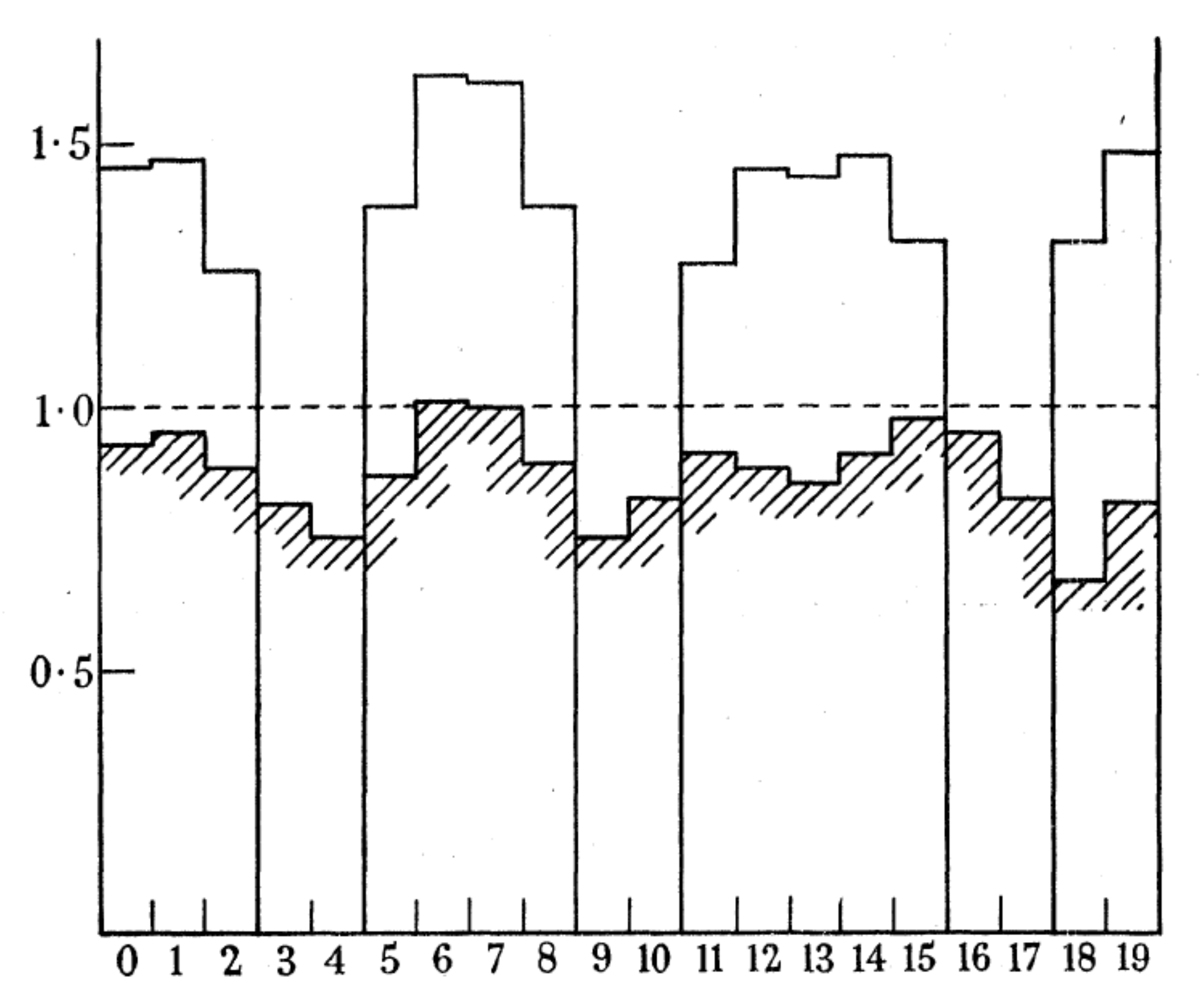}
\end{center}
\caption{The evolution of the ring after the bifurcation induced by a
diffusion-driven instability. The hatched graph represents the $X$
concentration at the initial state: all cells are, up to small fluctuations,
in the equilibrium state $\left( X_{0}=1,Y_{0}=1\right) $. The other graph
represents a stationary oscillatory wave pattern with $3$ lobes.\ The
divergences induced by the stability are tamed by two factors: (i) the
concentration $X$ cannot become $<0$, (ii) non-linear effects.}
\label{Turing_fig3}
\end{figure}

\section{Further aspects of Turing's paper}

In his paper, Turing evoked many other problems. In \S 4 he gave simple
examples for explaining the idea of \textquotedblleft breakdown of symmetry
and homogeneity\textquotedblright\ in pattern formation. He explained also
how exponential divergences are bounded by non-linearities which allow new
equilibria to emerge and he emphasized\ 

\begin{quotation}
\noindent \textquotedblleft the effect of considering non-linear reaction
rate functions when far from homogeneity.\textquotedblright ~(p.~58)
\end{quotation}

In \S 8 he listed some \textquotedblleft types of asymptotic behaviours in
the ring after a lapse of time\textquotedblright :

\begin{enumerate}
\item \textquotedblleft stationary cases\textquotedblright\ where the
asymptotic regime is dominated by a pair of \emph{real} eigenvalues $\left(
p_{s_{0}},p_{s_{0}}^{\prime }\right) $;

\item \textquotedblleft oscillatory cases\textquotedblright\ where the
asymptotic regime is dominated by a pair of \emph{complex conjugated}
eigenvalues $\left( p_{s_{0}},p_{s_{0}}^{\prime }\right) $ (travelling
waves);

\item \textquotedblleft limit cases\textquotedblright .
\end{enumerate}

He drew also \emph{phase diagrams} in the parameter space which classify
these different regimes and explained the role of fluctuations in the
bifurcation process.

\begin{figure}[ptb]
\begin{center}
\includegraphics[
width= 6.0912cm,height= 5.2697cm]{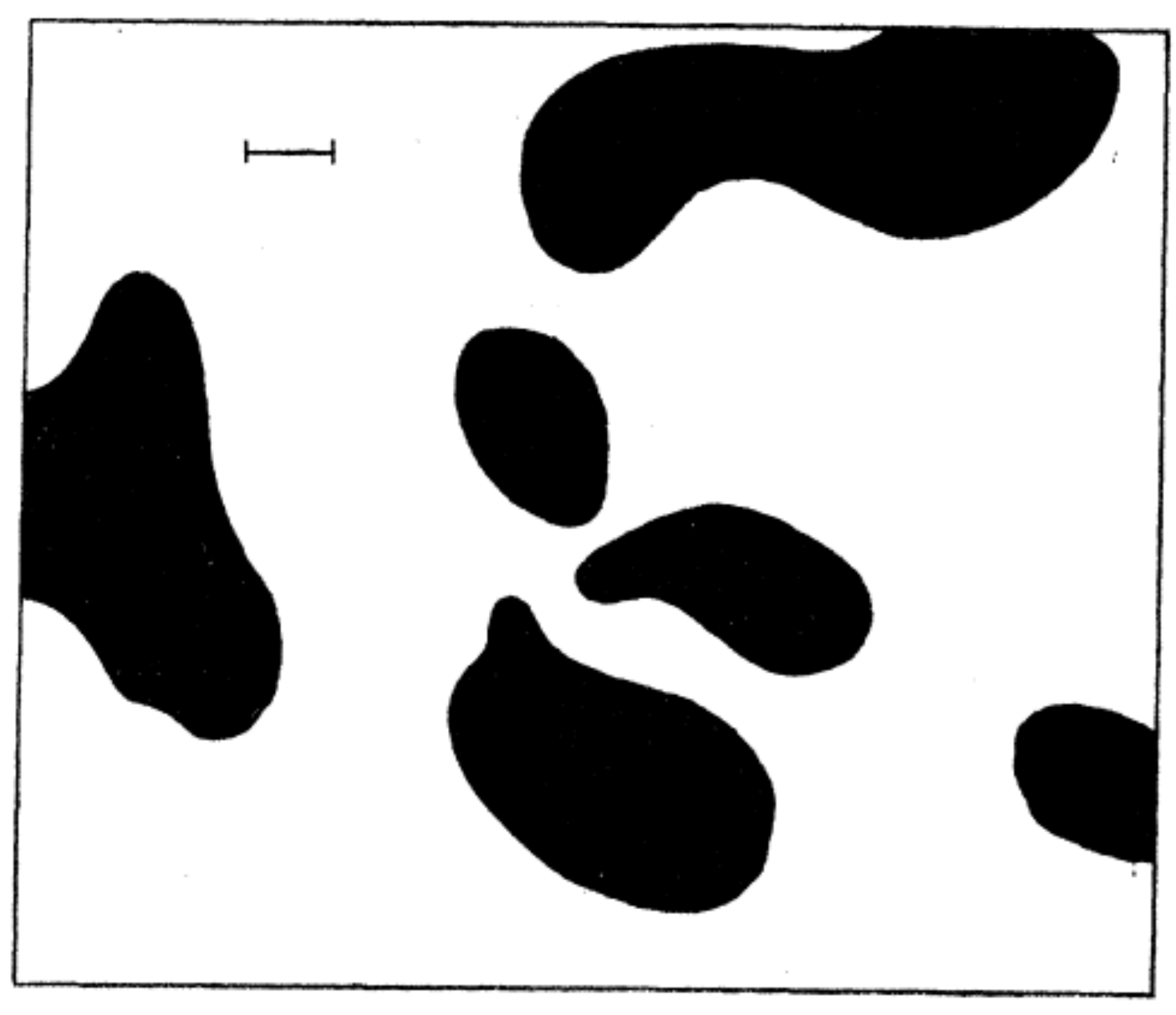}
\end{center}
\caption{Turing's figure 2 on \textquotedblleft dappled colour
patterns\textquotedblright\ in two-dimensionsional tissues.}
\label{Turing_fig2}
\end{figure}

Another extremely important anticipation is that in \emph{two-dimensionsional%
} tissues diffusion-driven instabilities can explain \textquotedblleft
dappled colour patterns\textquotedblright\ as observed on sea shells or
leopard's coats. Figure \ref{Turing_fig2} reproduces Turing's figure 2.



Moreover, Turing anticipated the fact that his general model was able to
induce \emph{oscillating} patterns when the chemical internal dynamics of
each cell bifurcates towards a \emph{limit cycle} by Hopf bifurcation. When
such limit cycles propagate spatially, many complex phenomena can emerge.\
Turing envisaged applications to organisms such as plants (flowers, leaves)
or Hydra. His predictions have been widely confirmed later.

In the fascinating \S 12 \textquotedblleft Chemical waves on spheres.\
Gastrulation\textquotedblright , Turing generalizes his one-dimensional ring
model to a \emph{two-dimensional sphere model} whose geometry is more
complex, the harmonic analysis on the sphere resting on the eigenfunctions
of the spherical Laplacian, namely the \emph{spherical harmonics}. His
striking idea was to apply the model to \emph{gastrulation} in embryology,
which is the step at which the spherical symmetry of the blastula is
broken.\ He developed this idea further in his unpublished paper on \emph{%
phyllotaxis}.

Finally, in the last \S 13 \textquotedblleft Non-linear theory.\ Use of
digital computer\textquotedblright , Turing came back to the assumption that
linearization is a good approximation and explained that it is the case only
in the neighbourhood of the first bifurcation.\ It is

\begin{quotation}
\noindent \textquotedblleft an assumption which is justifiable in the case
of a system just beginning to leave a homogeneous
condition.\textquotedblright ~(p.~72)
\end{quotation}

\noindent For Turing it was risky to try to go beyond:

\begin{quotation}
\noindent \textquotedblleft One cannot hope to have any very embracing
theory of such processes, beyond the statement of the
equations.\textquotedblright ~(p.~73)
\end{quotation}

\noindent Hence the fundamental interest of the just constructed digital
computers enabling numerical simulations avoiding the too drastic
simplifications imposed by the search of explicit theoretical solutions to
the equations.

\section{Conclusion: after Turing}

To conclude this presentation of Turing's 1952 paper, let us look briefly at
the works on reaction-diffusion equations after Turing. I have already
tackle this theme in my talk \cite{P03b} at the IAPS 2001 Conference on 
\emph{Complexity and Emergence}.

\subsection{General reaction-diffusion models}

Among the many specialists of the domain, we would cite Hans Meinhardt and
Alfred Gierer who, since 1972 \cite{Gierer}, have considerably increased
our knowledge on reaction-diffusion models. They have shown that, for an
activator/inhibitor pair of morphogens, instabilities mainly result from the
competition between a short range slow activation and a long range fast
inhibition, the inhibitor diffusing faster than the excitator. This confirms
Turing's remark on the role of the difference between the two diffusibility
coefficients $\mu $ and $\nu $.

A general Meinhardt-Gierer model has the form

\begin{equation*}
\left\{ 
\begin{array}{l}
\dot{x}=\rho \frac{x^{2}}{y}-\alpha x+\sigma _{x}+\mu \Delta x \\ 
\dot{y}=\rho x^{2}-\beta y+\sigma _{y}+\nu \Delta y%
\end{array}%
\right. ~\alpha <\beta ,~\mu \ll \nu
\end{equation*}

\noindent The activator morphogen $x$ is self-catalizing ($x^{2}$ term in $%
\dot{x}$) and its production is inhibited by the inhibitor morphogen $y$ ($%
\frac{1}{y}$ term in $\dot{x}$).\ Moreover, $x$ catalyzes its inhibitor ($%
x^{2}$ term in $\dot{y}$).\ The linear terms $\alpha x$ and $\beta y$ ($%
\alpha <\beta $) are degradation terms, the constant $\sigma _{y}$ enables a
stable homogeneous state and the constant $\sigma _{x}$ allows to trigger
the process. $\mu $ and $\nu $ are the diffusibility coefficients with $\mu $
(slow)$~\ll \nu $ (fast). A local fluctuation of the activator $x$ induces a
local peak of $x$ which diffuses slowly. But it amplifies also the inhibitor
concentration $y$, and since $y$ diffuses faster than $x$ it will inhibit
the production of $x$ at some distance (what is called a \textquotedblleft
lateral inhibition\textquotedblright ).

\begin{figure}[ptb]
\begin{center}
\includegraphics[
width= 12.1166cm,height= 6.5262cm]{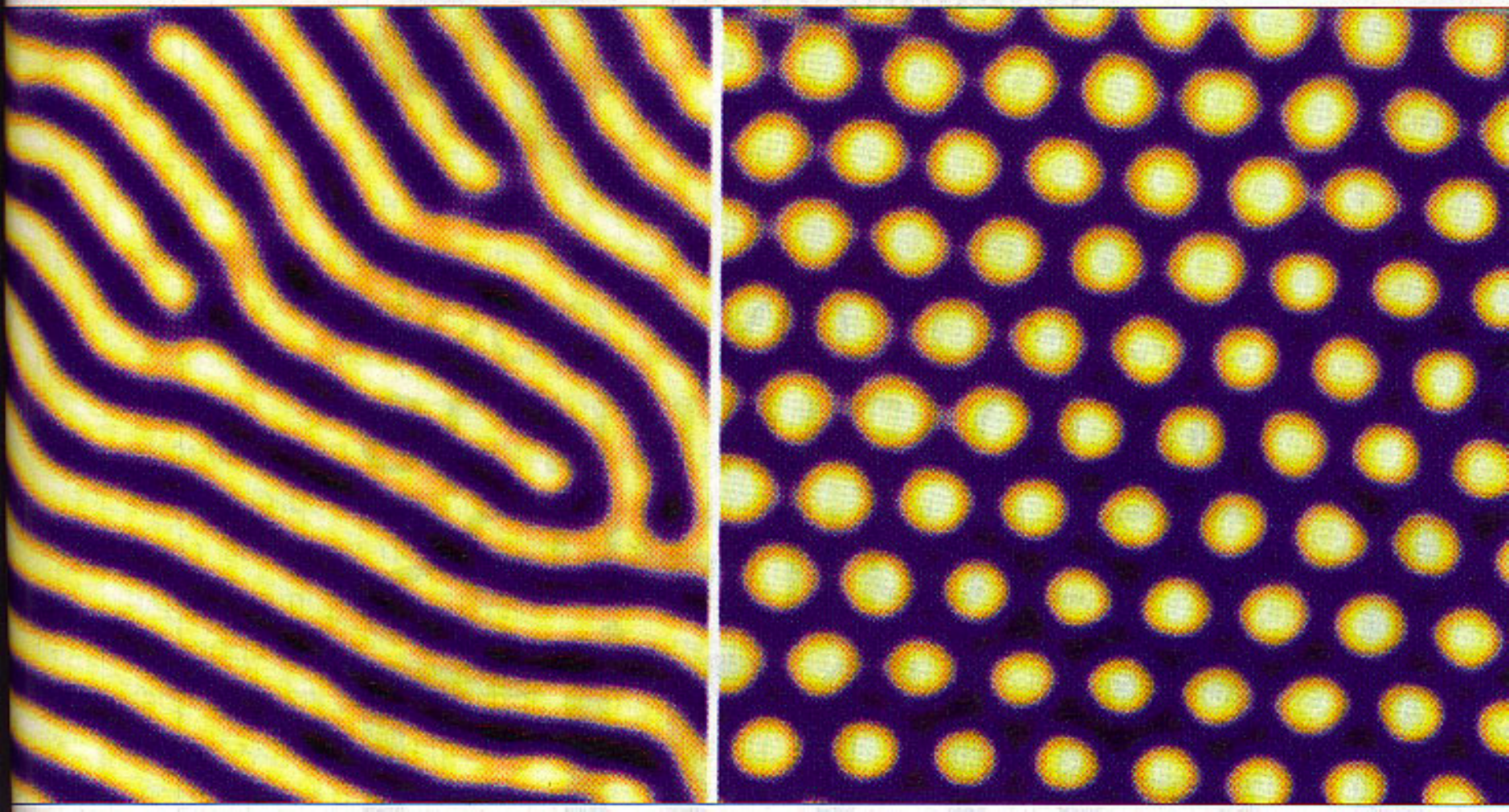}
\end{center}
\caption{The coupling between internal dynamics and external diffusion can
induce very complex patterns. Two examples due to De Kepper: a system of
stripes (with defects) and a honeycomb pattern. (From De Kepper \emph{et al.} \cite{Kepper}).}
\label{DeKepper}
\end{figure}

The coupling between the internal dynamics and external diffusion can induce
very complex patterns.\ Figure \ref{DeKepper} shows two examples due to De
Kepper \cite{Kepper}, a system of stripes with defects and a honeycomb pattern.



\begin{figure}[ptb]
\begin{center}
\includegraphics[
width= 10.1067cm,height= 7.3697cm]{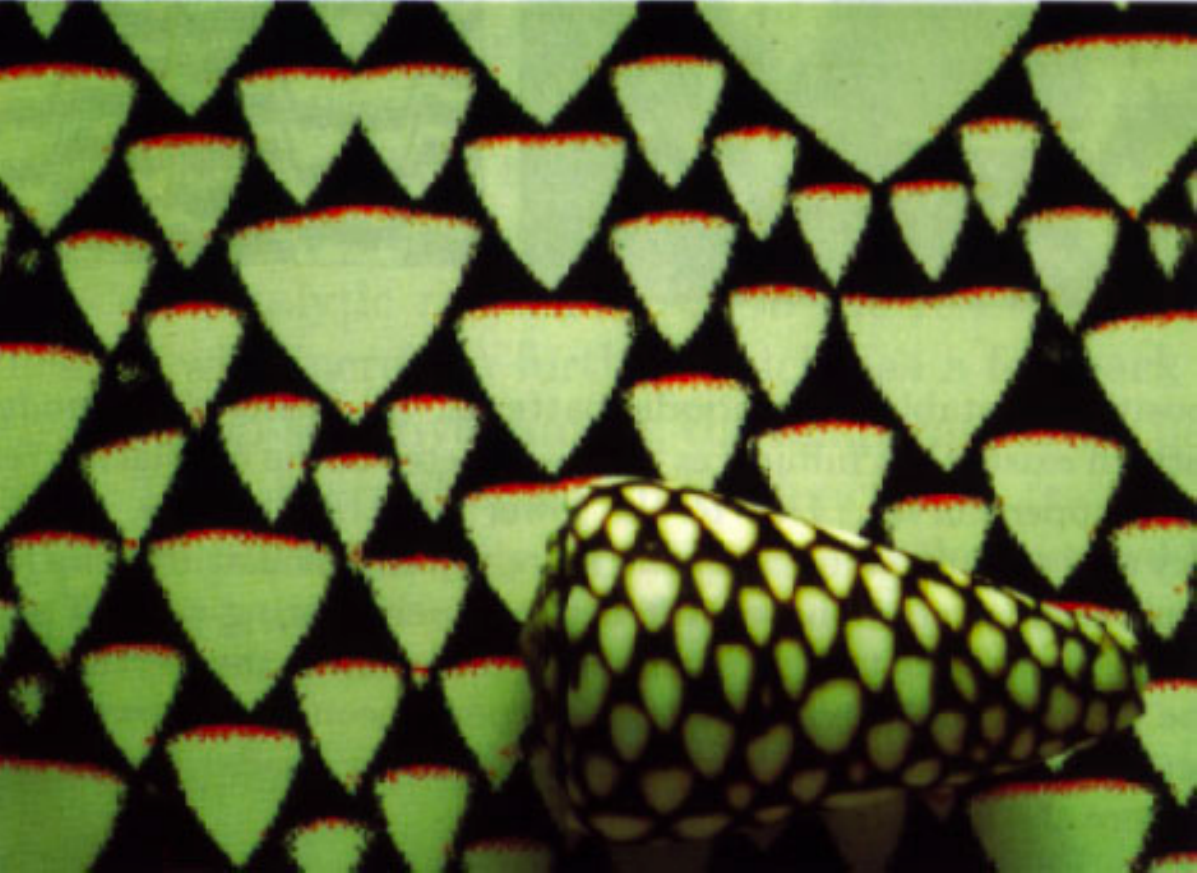}
\end{center}
\caption{Meinhardt's model for the sea shell \emph{Conus marmoreous}.\ In
front a true shell.\ In the background its reaction-diffusion model.\ (From
Meinhardt \protect\cite{Mein}).}
\label{Meinhardt_coq}
\end{figure}

One of the best known achievements of Hans Meinhardt is his modelling of sea
shells. Since the growth of a shell results from an accretion of calcified
matter along its boundary, it can be represented by a two-dimensional
diagram $B\times T=$ boundary$\times $time. The geometry is therefore in
fact one-dimensional. The diffusion of the activator from a local peak of
concentration induces a triangle where the pigmentation controlled by $x$ is
high, but the faster diffusion of the inhibitor stops it after a while.
Hence a cascade of triangles.\ Figure \ref{Meinhardt_coq} shows the
celebrated model of \emph{Conus marmoreous}.



\subsection{From Alan Turing to Ren\'{e} Thom}

As we have seen, in Turing's paper morphogenetic processes spatially unfold
diffusion-driven instabilities. In the late 1960s, Ren\'{e} Thom \cite%
{Thom72}, \cite{Thom74} proposed a more general model based on the general
concept of \emph{bifurcation}. The similarities and dissimilarities between
Turing's and Thom's models are fascinating. The key idea is the same:
internal chimical dynamics (reactions) are coupled with external spatial
dynamics, the latter destabilize the former and morphologies spatially
unfold the instabilities. As we have seen, in Turing coupling is given by
the diffusion of the reacting morphogens. In Thom, coupling is more
generally a spatial control of internal dynamics and the morphogenetic
discontinuities breaking the homogeneity of the substrate are induced by
bifurcations.

\subsection{Beyond Turing}

After Turing, many authors, e.g. James Murray \cite{Murray89}, \cite%
{Murray90}, introduced bifurcations in reaction-diffusion equations. Let $u$
be the vector $\left( x,y\right) $ and consider a differential equation $%
\dot{u}=f\left( u,r\right) $ where $r$ is a spatial control. When $r$ varies
and crosses a critical value $r_{c}$, the initial stable equilibrium state $%
u_{0}$ of the system can collapse with an unstable equilibrium and
disappear.\ The system is therefore projected to another equilibrium through
this \emph{saddle-node bifurcation}. Another most used bifurcation is the 
\emph{Hopf bifurcation}. When $r$ varies and crosses a critical value $r_{c}$%
, the initial stable (i.e. attracting) equilibrium state $u_{0}$ becomes a
repellor and generates a small attracting closed orbit (i.e. a limit cycle). 

Consider for instance the following system analyzed by Robin Engelhardt \cite%
{Engel}:

\begin{equation*}
\left\{ 
\begin{array}{l}
\dot{x}=-xy^{2}+ay-\left( 1+b\right) x+\delta \Delta x \\ 
\dot{y}=xy^{2}-\left( 1+a\right) y+x+F+\delta \Delta y%
\end{array}%
\right.
\end{equation*}

\noindent The chemical internal equilibria without diffusion ($\delta =0$)
are solutions of the equations (if $y^{2}+1+b\neq 0$):

\begin{gather*}
x=\frac{ay}{y^{2}+1+b} \\
a\frac{y^{3}}{y^{2}+1+b}-\left( 1+a\right) y+\frac{ay}{y^{2}+1+b}+F=0
\end{gather*}

\noindent that is

\begin{equation*}
y^{3}-Fy^{2}+\left( 1+b+ab\right) y-F\left( 1+b\right) =0
\end{equation*}

\noindent which is a cubic equation with parameters $a,b,F$.

\begin{figure}[tbp]
\begin{center}
\includegraphics[
width= 12.1166cm,height= 12.1166cm]{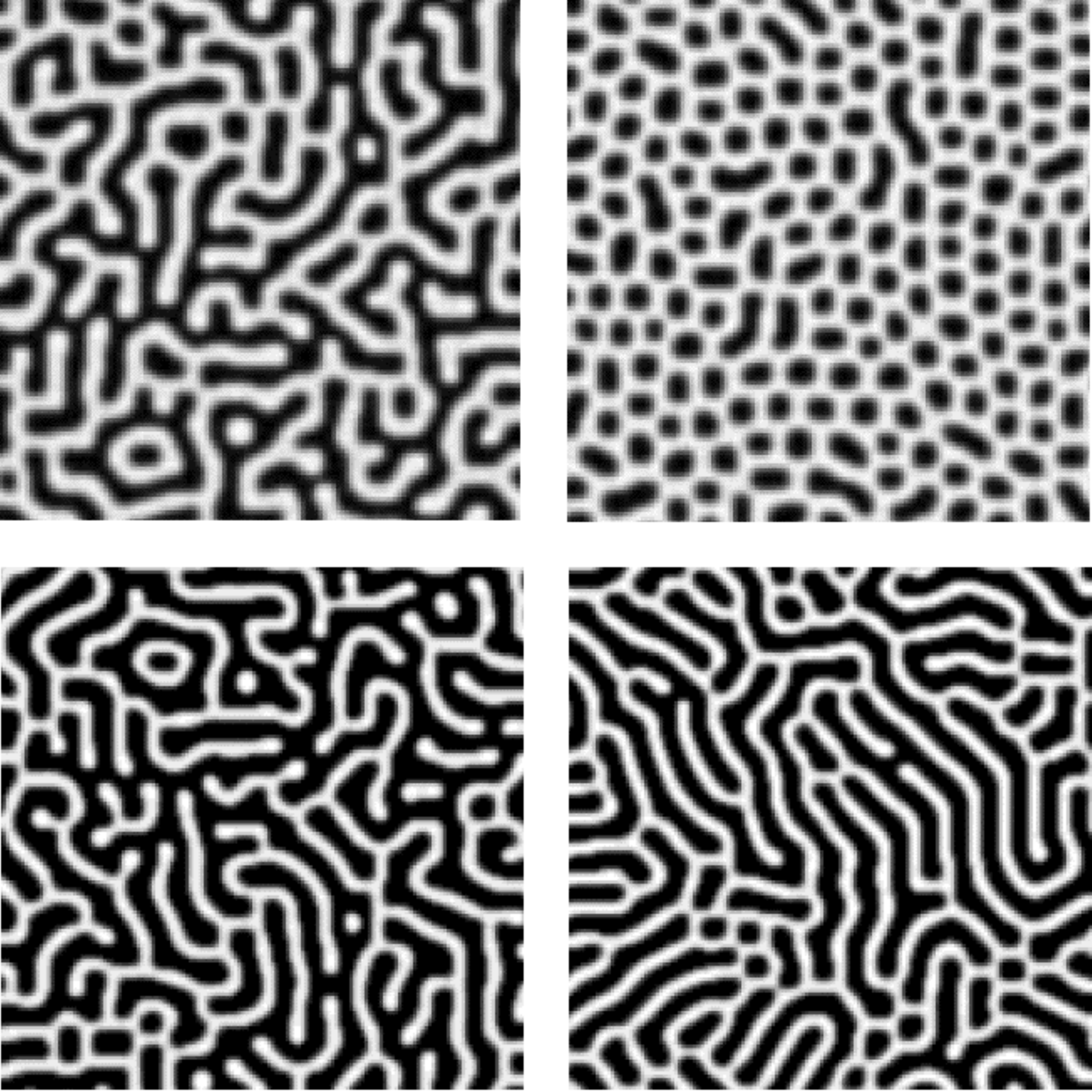}
\end{center}
\caption{Some examples of patternized solutions of Engelhardt's system of
equations. (From \protect\cite{Engel}). }
\label{EngelhardtTable}
\end{figure}

At the points where $y^{2}+1+b=0$, we have 
\begin{equation*}
\left\{ 
\begin{array}{l}
\dot{x}=ay+\delta \Delta x \\ 
\dot{y}=-bx-\left( 1+a\right) y+F+\delta \Delta y%
\end{array}%
\right.
\end{equation*}

\noindent and this can be an equilibrium point for $\delta =0$ only if $ay=0$
and $-bx-y+F=0$. If $a\neq 0$, this implies the condition $b=-1$, and the
equilibrium is $y=0$, $x=-F$. If $a=0$, the equilibrium would be $-bx-y+F=0$
with $y^{2}+1+b=0$. Figure \ref{EngelhardtTable} shows some examples of
patterns solution of this system of equations.



There is a wealth of material on these topics.\ The reader could look e.g.
at Harrison \cite{Harrison87}, Lee \emph{et al.} \cite{Lee1}, \cite{Lee2},
Maini \cite{Maini12}, Oyang-Swinney \cite{Oyang}, or Pearson \cite{Pearson}.

\subsection{Experimental results}

The validity of Turing's models for embryology are still under discussion.\
But in what concerns chimical and biological patterns their validity is
without doubt. We have seen Meinhardt's examples. For chemical systems exact
verifications go back to 1990 and the works of the Bordeaux group of Patrick
De Kepper (Castets, Dulos, Boissonade) on iodate-ferrocyanide-sulfite or
clorite-iodide-malonic acid-starch reactions in gel reactors.\ 

It is a full universe of morphological phenomena and mathematical models
that Turing opened in 1952 with a remarkable foresightedness.

\end{document}